%% file: main.tex
\DocumentMetadata{lang=en-US}
\documentclass{article} 
\usepackage{iclr2027_conference,times}

\input{math_commands.tex}

\def\mOmega{{\bm{\Omega}}}
\def\mGamma{{\bm{\Gamma}}}

\usepackage{todonotes}

\usepackage{amssymb}
\usepackage{booktabs}
\usepackage{dcolumn}
\newcolumntype{d}[1]{D{.}{.}{#1}}
\usepackage{siunitx}
\usepackage{mathtools}
\usepackage{nicefrac}
\usepackage{xcolor}
\definecolor{darkblue}{rgb}{0.0, 0.0, 0.45}
\usepackage[colorlinks,citecolor=darkblue]{hyperref}
\usepackage{cleveref}
\usepackage{url}

\title{Learning Low-Order Approximations of the\newline Quark Propagator in Lattice QCD}

\author{Simon Pfahler, Maren Käfferlein, Daniel Knüttel, Christoph Lehner, Tilo Wettig\\
Department of Physics\\
University of Regensburg\\
93040 Regensburg \\
\texttt{simon.pfahler@ur.de} \\
}

\begin{document}

\maketitle

\begin{abstract}
	The hopping expansion is an important approximation of the quark propagator in lattice quantum chromodynamics, but it diverges for light quarks, which limits the scope of strong-coupling and finite-density simulations.
	We present an artificial scientific discovery framework using gauge-equivariant neural networks in the Clifford basis to extract interpretable approximations.
	Our approach yields improved low-order approximations that outperform the hopping expansion at identical computational cost.
	In the parameter regime where the hopping expansion fails, our approach identifies convergent behavior, thus suggesting a novel series expansion.
	The discovered approximations generalize to unseen gauge fields and volumes, enabling immediate implementation as plug-in replacements and a new theoretical understanding of quark-propagator expansions.
\end{abstract}

\section{Introduction}
Lattice quantum chromodynamics (QCD) is currently the most precise non-perturbative method to compute theoretical predictions for many particle-physics processes~\citep{FlavourLatticeAveragingGroupFLAG:2024oxs}.
These predictions also enable the search for new physics beyond the Standard Model, including the nature of dark matter~\citep{Kronfeld:2012uk}, the muon anomalous magnetic moment~\citep{Aliberti:2025bega}, or the origin of neutrino masses~\citep{Davoudi:2024ukx}.
However, obtaining these predictions is computationally very expensive, using roughly \SI{20}{\percent} of the compute available at scientific data centers~\citep{Suarez:2025cnd,Camps:2025adg}.

In lattice QCD space-time is discretized to enable computer simulations~\citep{Wilson:1974sk}.
An important element of these simulations is the calculation of propagators, which describe how quarks propagate from one space-time point to another.
Evaluating these propagators amounts to solving the Dirac equation, which on the discretized space-time lattice becomes a large, sparse linear system of equations.
The corresponding matrix is called the Dirac operator or matrix.

A prominent method to obtain approximate solutions of the Dirac equation is the hopping expansion~\citep{Hasenfratz:1981nc}, which is a power-series expansion with a finite radius of convergence.
Because of its simplicity and practicality, the hopping expansion is widely used in lattice QCD, e.g., as an algorithmic tool for error reduction in Hybrid Monte Carlo simulations~\citep{Hasenbusch:2018iuw}, to enable heavy dynamical quark simulations and calculations in the strong-coupling regime~\citep{Wakabayashi:2021eye,Philipsen:2021vgp,Langelage:2010yn,Nakamura:2015jra}, to address sign problems at nonzero baryon chemical potential~\citep{Goy:2016egl}, and to power reweighting methods for finite-density simulations~\citep{Barbour:1988ax,Bornyakov:2016wld}.

While the hopping expansion is the optimal series expansion inside its radius of convergence, low-order truncations may not result in optimal approximations of the quark propagator.
For the physically interesting case of light quarks the expansion diverges~\citep{Wakabayashi:2021eye}, which currently restricts strong-coupling effective theories~\citep{Philipsen:2021vgp} and finite-density methods~\citep{Goy:2016egl} to heavy quarks.

In this work, we investigate improvements to low-order truncations of the hopping expansion using machine learning, specifically through the lens of neurosymbolic AI~\citep{Bhuyan_2024}.
Rather than using neural networks as black-box predictors, we employ them as discovery engines that identify improved approximations with interpretable coefficients.
Our architecture is grounded in geometric deep learning~\citep{Bronstein:2021mdi}, using gauge-equivariant neural networks that respect the fundamental $\textrm{SU}(3)$ gauge symmetry of QCD by construction~\citep{Favoni:2020reg}, in analogy to how group-equivariant networks in computer vision respect rotational and translational invariance~\citep{Cohen_2016}.
The main contributions of our work are:
\begin{enumerate}
	\item[(1)] \textbf{Artificial Scientific Discovery Framework.}
	      We introduce neural-network architectures that operate directly in the Clifford basis of $\mathbb C^{4\times4}$ in order to obtain model weights that are directly interpretable.
	      From the model weights, we extract expansion coefficients as a plug-in replacement for applications that currently use the hopping expansion.
	\item[(2)] \textbf{Improved low-order approximations.}
	      Within the convergence radius of the hopping expansion, we find improved low-order approximations of the quark propagator that outperform the hopping expansion at the same computational cost.
	\item[(3)] \textbf{Evidence of a novel expansion.}
	      Outside the hopping expansion's convergence radius, our models discover convergent approximations, indicating a novel series expansion.
	\item[(4)] \textbf{Transferability.}
	      We show that our improved approximations generalize perfectly (within statistics) to unseen gauge fields drawn from the same probability distribution and to a different lattice volume, indicating that they capture general properties of the Wilson Dirac operator.
	      We also show good transferability to gauge fields following a different probability distribution, in particular in the regime where the hopping expansion fails to converge.
\end{enumerate}

\section{Related Work}
Our work lies at the intersection of gauge-equivariant neural networks for lattice QCD, machine-learning approaches to the Dirac equation, and artificial scientific discovery via symbolic regression.

\textbf{Gauge-equivariant neural networks.}
Neural networks that respect group symmetries were first introduced in \citet{Cohen_2016} for generic groups and adapted to lattice gauge theories in \citet{Favoni:2020reg}, who introduced lattice gauge-equivariant convolutional neural networks.
\citet{Lehner:2023bba,Lehner:2023prf} showed that gauge-equivariant neural networks can learn preconditioners for the Wilson-clover Dirac operator, requiring minimal retraining across gauge configurations of the same ensemble.
Our work builds directly on the gauge-equivariant architecture of \citet{Pfahler:2026fpj}, which also added long-range interactions to improve the preconditioner near the critical mass.
While these works focus on accelerating iterative solvers, we use a similar architecture to learn improved low-order expansions of the quark propagator.

\textbf{Machine learning for lattice QCD.}
Beyond preconditioning, machine learning has been applied to several aspects of lattice QCD, including flow-based sampling to mitigate critical slowing down~\citep{Kanwar:2020xzo} and variance reduction for correlation functions~\citep{Abbott:2026ylv}.
A recent review is given by \citet{Boyda:2022nmh}.
\citet{Zhang:2026bqa} used machine learning to discover analytical formulas from data obtained via lattice QCD.
However, to our knowledge, no prior work uses machine learning to discover low-order approximations of the quark propagator.

\textbf{Symbolic regression and artificial scientific discovery.}
Our work relies heavily on symbolic regression~\citep{Dong_2025}, in which a variety of approaches is being used, including deep-learning methods~\citep{Petersen_2021}, transformers~\citep{Biggio_2020}, genetic programming~\citep{Mundhenk_2021}, and equation learners~\citep{Martius_2016}.
Since much of the structure of the expansions is known a priori due to underlying symmetries, we use the equation-learner approach introduced by \citet{Martius_2016} and adapt it to include the mathematical operations required in our application.
In physics, finding formulas that fit noisy data is a central problem, and many efforts have been made to automate this task.
While early work mostly focused on genetic-programming approaches, such as Eureqa~\citep{Schmidt_2009} or AI Feynman~\citep{Udrescu:2019mnk}, many more approaches are used nowadays, see \citet{Wetzel:2025uhj} for a review.
A related recent work by \citet{Kitazawa:2026vbv} uses human-AI collaboration to provide more efficient algorithmic implementations of higher-order truncations of the hopping expansion, but does not tune coefficients.

\section{Problem Formulation}
\subsection{Dirac Operator and Hopping Expansion}\label{sec:QCD_background}
In lattice QCD, continuous space-time is discretized on a space-time lattice, turning quantum field theoretical problems into numerically tractable quantum mechanics problems.
One fundamental object on the lattice is the quark field, which lives on the lattice sites and typically has $4$ spin and $3$ color degrees of freedom per lattice site.
Another object is the gauge field, which connects neighboring lattice sites and acts on the color degrees of freedom of the quark field.
On each link on the lattice, the gauge field is represented by a matrix in $\textrm{SU}(3)$, the group of complex three-dimensional matrices with determinant equal to 1.

We consider a four-dimensional lattice with $L_x,L_y,L_z,L_t$ lattice sites.
On this lattice, we define quark fields $\va\in\sC^{V\times4\times3}$, where $V=L_x\times L_y\times L_z\times L_t$, and gauge fields $\mU_\mu\in\textup{SU}(3)^{V}$, where $\mu=1,2,3,4$ is a space-time dimension.
We use the shorthand notation $\vx$ for the space-time indices $x,y,z,t$.
We further define gauge transformations $\mOmega\in\textup{SU}(3)^V$ that act on the color degrees of freedom and transform the quark and gauge fields according to
\begin{equation}\label{eq:gauge_transformation}
	\va(\vx,s)\rightarrow\mOmega(\vx)\va(\vx,s)\,,\qquad\mU_\mu(\vx)\rightarrow\mOmega(\vx)\mU_\mu(\vx)\mOmega^\dag(\vx+\ve^{(\mu)})\,.
\end{equation}
Here, $s$ is a spin index and $\ve^{(\mu)}$ is the unit-vector in the $\mu$-direction.
QCD obeys a gauge symmetry, which implies that physical observables (constructed from $\va$ and $\mU$) remain invariant under \cref{eq:gauge_transformation}.
This will become important when we construct our neural networks in \cref{sec:Architecture} below.

Linear combinations such as $\va(\vx)+\va(\vx+\ve^{(\mu)})$ are incompatible with this gauge symmetry.
In order to allow for information transfer between neighboring lattice sites while respecting the gauge symmetry, we introduce the hop operator $\mH_{\pm\mu}\in\sC^{V\times3\times3}$ through its action on a quark field $\va$,
\begin{subequations}\label{eq:hop_operator}
	\begin{align}
		\mH_{+\mu}\va(\vx,s) & =\mU^\dag_\mu(\vx-\ve^{(\mu)})\va(\vx-\ve^{(\mu)},s)\,, \\
		\mH_{-\mu}\va(\vx,s) & =\mU_\mu(\vx)\va(\vx+\ve^{(\mu)},s)\,.
	\end{align}
\end{subequations}
Under a gauge transformation, $\mH_{\pm\mu}\va(\vx,s)\rightarrow\mOmega(\vx)\mH_{\pm\mu}\va(\vx,s)$.
This is known as gauge equi\-variance.
Linear combinations such as $\va(\vx)+\mH_{\pm\mu}\va(\vx)$ are compatible with the gauge symmetry.

By applying multiple hop operators in sequence, we can transport information over longer distances.
We denote a path $p$ by a list of pairs, where each pair $(\pm\mu, d)$ consists of the space-time direction $\pm\mu$ and the distance $d\in\sN$ to indicate a transport in direction $\pm\mu$ by a distance of $d$.
A given pair $h=(\pm\mu, d)$ translates to an operator $\mH_h=(\mH_{\pm\mu})^{d}$.
The parallel-transport operator $\mT_p$ that applies a transport along a path $p=[h_1,\ldots,h_n]$ can then be written as
\begin{equation}
	\mT_p=\mH_{h_n}\ldots\mH_{h_1}\,.
\end{equation}

The Wilson Dirac equation is the linear equation $\mD\va=\vb$, where $\mD$ is the Wilson Dirac operator introduced by~\citet{Wilson:1974sk},
\begin{equation}\label{eq:Dirac_operator}
	\mD=(m+4)\mI-\frac12\sum_{\mu=1}^4\left((\mI_4+\gamma_\mu)\mH_{+\mu}+(\mI_4-\gamma_\mu)\mH_{-\mu}\right).
\end{equation}
Here, $m\in\sR$ is the mass parameter, $\gamma_\mu\in\sC^{4\times4}$ are the Euclidean gamma matrices satisfying the anticommutation relation $\gamma_\mu\gamma_\nu+\gamma_\nu\gamma_\mu=2\delta_{\mu\nu}$, and $\mI_4,\gamma_\mu$ act on the spin degrees of freedom.

Note that the mass parameter $m$ in \cref{eq:Dirac_operator} is not the physical quark mass, which is only recovered after renormalization.
Light quark masses (for up, down, and strange quark) correspond to mass parameters $m<0$.

In order to approximate solutions to the Wilson Dirac equation, we need approximations to $\mD^{-1}$.
In this work, we consider the hopping expansion.
We rewrite the Wilson Dirac operator in the form
\begin{equation}\label{eq:Dirac_operator_2}
	\mD
	=\frac1{2\kappa}\bigg(\mI-\kappa \sum_{\mu=1}^4\big((\mI_4+\gamma_\mu)\mH_{+\mu}+(\mI_4-\gamma_\mu)\mH_{-\mu}\big)\bigg)
	=\frac1{2\kappa}\big(\mI-\kappa \mH\big)
\end{equation}
with $\kappa=\frac1{2(m+4)}$ and the hopping term $\mH$.
Since $\mH_{\pm\mu}$ is unitary, its spectral norm is $1$.
So is the spectral norm of $\gamma_\mu$.
The spectral norm of $\mH$ is therefore bounded by $8$.
Thus, there is a critical $\kappa_h>\nicefrac18$ (or $m_h<0$) such that for $\kappa<\kappa_h$ (or equivalently $m>m_h$), $\mD^{-1}$ can be expanded as a series in powers of $\kappa$, resulting in the hopping expansion~\citep{Hasenfratz:1981nc}
\begin{equation}\label{eq:Hopping_expansion}
	\mD^{-1}=2\kappa\sum_{\ell=0}^\infty\kappa^\ell\mH^\ell\,.
\end{equation}

In applications, the series has to be truncated at a given $\ell=n$.
With this truncation, it is unclear whether the hopping expansion is still the optimal approximation or if there are other approximations that require the same computational cost but provide a better approximation of $\mD^{-1}$.

Furthermore, it is unclear whether a different series expansion of the quark propagator can be found in the regime $m<m_h$, where the hopping expansion does not converge anymore~\citep{Wakabayashi:2021eye}.
Strong-coupling effective theories~\citep{Philipsen:2021vgp} and finite-density methods~\citep{Goy:2016egl} are currently restricted to heavy quarks because of this convergence limit.

\subsection{Objective}
In order to investigate the limitations and possible improvements of truncated hopping expansions, we quantify how well an operator $\mM$ approximates the inverse of the Dirac operator $\mD$ by the error
\begin{equation}
	\varepsilon(\mM,\mD)=\frac{\|\mM\mD-\mI\|_F^2}{\|\mI\|_F^2}\,,
\end{equation}
where $\|\bullet\|_F$ denotes the Frobenius norm, and the denominator is introduced to remove the dependence on the lattice volume.
Evaluating the Frobenius norm is expensive since the dimension of the matrices involved is $N=L_xL_yL_zL_t\cdot4\cdot3$.
We therefore rewrite the Frobenius norm as an expectation value over random normally distributed and normalized quark fields $\rvv$ and obtain
\begin{equation}\label{eq:Approximation_quality}
	\varepsilon(\mM,\mD)=\E\left(\|\mM\mD\rvv-\rvv\|^2\right),
\end{equation}
see \cref{app:Approximation_quality}, with $\|\bullet\|$ denoting the vector norm.

In this work, we train machine-learning models to learn operators $\mM$ with minimal $\varepsilon(\mM,\mD)$ under a fixed computational budget to search for improved low-order quark propagator approximations.

\section{Methods}
\subsection{Gauge-Equivariant Neural Networks}\label{sec:Architecture}
The Dirac operator transforms under gauge transformations as $\mD(\vx,\vy)\rightarrow\mOmega(\vx)\mD(\vx,\vy)\mOmega^\dag(\vy)$, and the same needs to be true for its inverse.
We have seen that linear combinations of hop operators are compatible with the gauge symmetry, so we can build approximations to $\mD^{-1}$ from such linear combinations.
A model built in this way does not need to learn the gauge symmetry, which is already baked into its architecture.
Therefore, the amount of training needed is significantly reduced.

We introduce two types of layers for our models, Clifford layer (C) and hopping layer (H).
Clifford layers take $n$ quark fields $\va_j$ and construct $m$ linear combinations $\vb_i$ from them,
\begin{equation}\label{eq:Clifford_layer}
	\vb_i(\vx)\overset{\textrm C}=\sum_{j=1}^n\sum_{k=1}^{16}w_{ijk}\mGamma_k\va_j(\vx)\quad\text{for }i=1,\ldots,m\,,
\end{equation}
where the $w_{ijk}\in\sC$ are trainable layer weights and $\mGamma_k$ is the Clifford basis of $\sC^{4\times4}$.\
\footnote{The Clifford basis consists of $\mI_4$, $\gamma_\mu$, $\frac i2[\gamma_\mu,\gamma_\nu]$ for $\mu<\nu$, $i\gamma_\mu\gamma_5$, and $\gamma_5$ with $\gamma_5=\gamma_1\gamma_2\gamma_3\gamma_4$ and $i$ the imaginary unit.}
This parametrization allows for the direct interpretation of the model weights and leads to faster training, as the model does not have to learn the structure of the $\mGamma_k$, see \cref{app:Clifford_vs_4x4}.

\begin{figure}[ht]
	\centering
	\includegraphics[height=23ex]{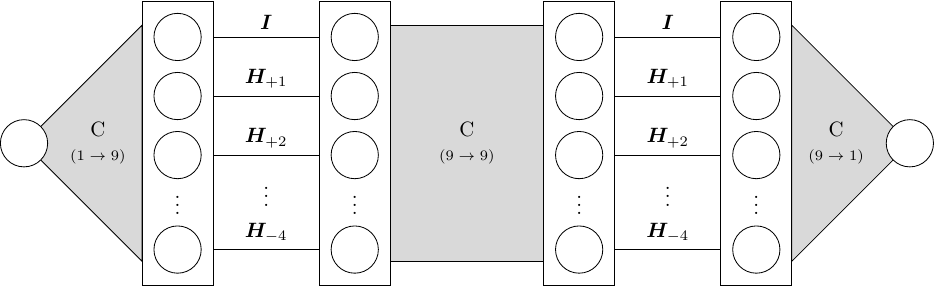}
	\vspace{-1ex}\caption{Sketch of the architecture of an HC model. Circles represent quark fields, gray areas represent C layers, and horizontal lines represent the different operations inside the H layer. To construct deeper models, more pairs of H and C($9\rightarrow9$) layers can be added in the middle.}
	\label{fig:Architecture}
\end{figure}

The hopping layer is a special case of the parallel-transport layer in \citet{Pfahler:2026fpj}.
It takes as input $9$ quark fields $\va_i$ and applies hop operators to them to construct $9$ output quark fields $\vb_i$,
\begin{equation}\label{eq:Hopping_layer}
	\vb_i(\vx)\overset{\textrm H}=
	\begin{cases}
		\va_i(\vx)\,,             & i=1\,,       \\
		\mH_{+(i-1)}\va_i(\vx)\,, & i=2,3,4,5\,, \\
		\mH_{-(i-5)}\va_i(\vx)\,, & i=6,7,8,9\,.
	\end{cases}
\end{equation}
Here, $i=1$ corresponds to the identity, $i=2,3,4,5$ correspond to hops in the forward direction along every dimension, and $i=6,7,8,9$ correspond to backward hops.
By alternating H and C layers, we can build arbitrarily deep neural networks, see \cref{fig:Architecture}.
We call a model built from these layers an HC model.
Such a model applies the same set of hop operators per H layer as one step of the hopping expansion.
Since the hop operations are the computationally expensive part of this calculation,\footnote{Lattice QCD calculations are typically performed by parallelizing over the lattice. Hop operations need communication between the parallel processes, which is why they dominate runtime.} applying an $\ell$-layer HC model\footnote{When referring to the number of layers of an HC model, we mean the number of H layers in the model.} has the same computational cost as applying an $\ell$-step truncated hopping expansion.
Note that both the hopping expansion and polynomials in $\mD$ can be realized by appropriate choices of the model weights.

\subsection{Restricted Model}\label{sec:Restricted}
We also introduce a second model architecture that is more closely related to the hopping expansion.
The fundamental layer used for this model is the restricted hopping layer (R), adapted from the hopping term $\mH$ in \cref{eq:Dirac_operator_2}.
It takes as input a single quark field $\va$ and outputs a single quark field $\vb$,
\begin{equation}
	\vb(\vx)\overset{\textrm R}=\sum_{\mu=1}^4\big((w_\mI\mI_4+w_\gamma\gamma_\mu)\mH_{+\mu}+(w_\mI\mI_4-w_\gamma\gamma_\mu)\mH_{-\mu}\big)\va(\vx)\,,
\end{equation}
with two trainable weights $w_\mI$ and $w_\gamma$.
A restricted model with $n$ restricted hopping layers is defined similarly to how the hopping expansion is defined via the hopping term,
\begin{equation}
	\mM=w\sum_{\ell=0}^n\prod_{\ell'=1}^\ell \textrm{R}_{\ell'}\,,
\end{equation}
see \cref{eq:Hopping_expansion}.
Note that we allow for different weights in each of the R layers, and that there is one additional trainable weight that scales the entire formula, similar to the $2\kappa$ factor in \cref{eq:Hopping_expansion}.

The HC model is a generalization of this restricted model.
While the HC model is more expressive, the restricted model is easier to train, and its weights are easier to analyze.
We will see the advantages of both models in \cref{sec:Experiments}.

\subsection{Training Setup}\label{sec:Training_setup}
The Wilson Dirac operator $\mD$ depends on the gauge field $\mU$ and the mass parameter $m$.
Physical observables are obtained as an average over gauge configurations for fixed $m$.
In training, we therefore keep the mass fixed and use a number $N_\mU$ of different gauge fields to avoid overfitting to a single gauge field.
Our loss function is
\begin{equation}\label{eq:Loss}
	C(\mM, m)=\frac1{N_\mU}\sum_{i=1}^{N_\mU}\|\mM_{\mU_i}\mD_{\mU_i,m}\rvv_i-\rvv_i\|^2
\end{equation}
for random normally distributed and normalized $\rvv_i$, see \cref{eq:Approximation_quality}.
The subscripts on $\mM$ and $\mD$ indicate the gauge field and mass parameter for which these operators are evaluated.
In practice, we used $N_\mU=4$ and obtained agreement between training loss and test loss within statistics.

Preliminary measurements indicated that the expectation value in \cref{eq:Approximation_quality} is sufficiently stable, with a worst observed coefficient of variation of \SI{1.63}{\percent}.
This justifies the use of a single sample\footnote{On the lattice volume $8^3\times16$ used in training, a single sample of $\rvv$ contains \num{196607} random numbers.} $\rvv$ in \cref{eq:Loss} per gauge configuration and training step.
The gauge fields used for training are taken from the ensemble in \citet{Knuttel_2025}, which is an ensemble of gauge configurations on an $8^3\times16$ lattice, created in the quenched approximation and using the Wilson action at coupling parameter $\beta=6$.

We trained the models using the Adam optimizer~\citep{Kingma:2014vow} with standard hyperparameters, i.e., learning rate $10^{-3}$, $\beta_1=0.9$, and $\beta_2=0.999$.
Every $100$ training iterations, we calculate a test score on gauge fields not used for training, and terminate when the test score changes by less than \SI{0.1}{\percent}, or after a maximum of \num{10000} training steps.
We initialize the model weights such that the initial model acts as an identity for $m<-0.5$, and such that it acts as the hopping expansion for $m>-0.5$.
In both cases, we add small complex random noise with standard deviation $10^{-3}$ to all model parameters.
This weight initialization led to stable training for all mass-parameter values.

\subsection{Weight Analysis and Scientific Discovery}\label{sec:Weight_analysis}
After training a model, we perform downstream analysis of the model weights to compare our models and the hopping expansion.
To this end, we rewrite the model in terms of coefficients $c_{p,k}$ of parallel transports along different paths, i.e.,
\begin{equation}\label{eq:Weight_analysis}
	\mM=\sum_{\textrm{Paths } p}\sum_{k=1}^{16}c_{p,k}\mGamma_k\mT_{p}\,.
\end{equation}
The hopping expansion can be expressed in the same form with some $c_{p,k}=0$, which allows for a direct comparison of the model and the hopping expansion.
If the model learns a $c_{p,k}\ne0$ that is zero in the hopping expansion, it has discovered a new functional form beyond the hopping expansion.

In principle, we could have started with a model of the form \cref{eq:Weight_analysis} and fitted the $c_{p,k}$ via linear regression.
However, in this case the number of paths, and thus the number of $c_{p,k}$ to be fitted, grows exponentially with the number of layers, rendering this approach infeasible for many layers.

\section{Experiments}\label{sec:Experiments}
\subsection{Setup}
We train both restricted and HC models using the training setup described in \cref{sec:Training_setup}.
In order to get a good overview of the model behavior in different parameter regimes, we train models for 36 masses between $-5$ and $2$ in steps of $0.2$.
To assess the convergence behavior of the low-order approximations obtained using our models, we train models with different numbers of layers.

For testing, we evaluate the approximation quality of \cref{eq:Approximation_quality} on four gauge fields not used in training on each of two different lattice volumes, $8^3\times16$ and $16^3\times32$, both generated as described above.
Tests using different probability distributions for the gauge fields\footnote{\label{foot:distributions}Specifically, we considered dynamical (or unquenched) gauge fields and Haar-distributed gauge fields.} were also done, see \cref{app:Action_dependence}.
The $8^3\times16$ gauge fields are taken from \citet{Knuttel_2025}, and the $16^3\times32$ gauge fields from \citet{Knuttel_2025a}.
For each gauge field, we use $100$ random quark fields to evaluate $\varepsilon(\mM,\mD)$ and report mean and standard deviation across all measurements.

\subsection{Approximation Quality}\label{sec:Approximation_quality}
\begin{figure}[ht]
	\centering
	\includegraphics[height=24ex]{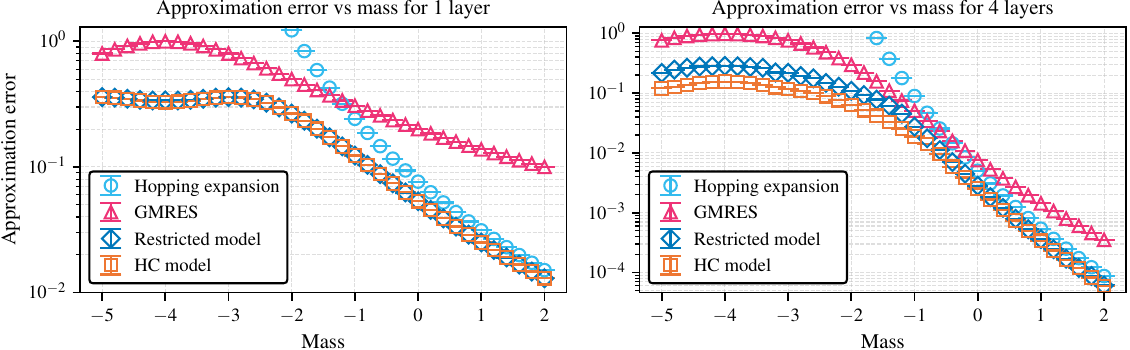}
	\vspace{-1ex}\caption{Approximation error of 1-layer models compared to 1 step of the hopping expansion and GMRES (left), and of 4-layer models compared to 4 steps of the hopping expansion and GMRES (right). The models systematically outperform the hopping expansion and GMRES across all mass parameters at the same computational cost (determined by the number of hop operations).}
	\label{fig:Approximation_quality}
\end{figure}

We first investigate the approximation quality of the models and compare it to the approximation quality of the hopping expansion, and to that of GMRES~\citep{Saad_2003}.
To obtain an approximation of $\mD^{-1}\va$ via GMRES, we solve $\mD\vx=\va$ with $n$ steps using $\bm0$ as initial guess, where $n$ matches the number of layers of the models and the order of the hopping expansion.
As GMRES adaptively builds an optimal polynomial approximation of $\mD^{-1}$ for each right-hand side $\va$, it provides a lower bound on the approximation error achievable by any fixed polynomial in $\mD$.
\Cref{fig:Approximation_quality} shows the approximation quality for 1- and 4-layer models and compares it to 1 and 4 steps of both the hopping expansion and GMRES.

In \cref{fig:Approximation_quality} we only show the larger $16^3\times32$ volume.
In fact, the approximation quality of our models is independent of the lattice volume within statistics (see \cref{app:Volume_dependence}), even though all training was done on the smaller $8^3\times16$ volume.
This is because our model architecture is similar to convolutional architectures and was designed to allow for transfer to larger volumes.
An investigation of the dependence on the probability distribution of the gauge fields is provided in \cref{app:Action_dependence}.

For all mass parameters, the models systematically outperform the hopping expansion as well as GMRES, with a mass-averaged reduction in approximation error of \SI{43}{\percent} (\SI{55}{\percent} for $m<m_h$, \SI{23}{\percent} for $m>m_h$) for 1-layer models compared to the best baseline (hopping expansion or GMRES).
For 4-layer models, the improvement increases to \SI{65}{\percent} ($m<m_h$: \SI{78}{\percent}, $m>m_h$: \SI{41}{\percent}) for HC models and to \SI{54}{\percent} ($m<m_h$: \SI{64}{\percent}, $m>m_h$: \SI{37}{\percent}) for restricted models.
This is because the models are a generalization both of the hopping expansion and of polynomials in $\mD$, and this additional freedom enables better approximations of $\mD^{-1}$.
In the regime $m>m_h$ where the hopping expansion is valid, the models learn the same structure as the hopping expansion (and thus HC and restricted models perform similarly here), but the additional freedom of layer-dependent weights leads to better low-order approximations.
This tuning is partially dependent on the underlying probability distribution of the gauge fields, so a new model should be trained when switching to gauge fields generated with a significantly different action (for details see \cref{app:Action_dependence}).

\begin{figure}[ht]
	\centering
	\includegraphics[height=24ex]{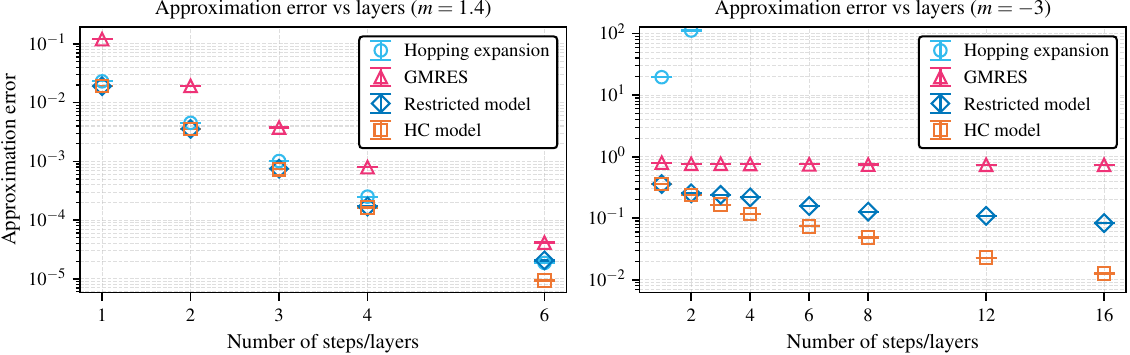}
	\vspace{-1ex}\caption{Dependence of approximation error on the number of layers (for models) or steps (for the hopping expansion and GMRES), for mass parameters $m=1.4$ (left) and $m=-3$ (right).}
	\label{fig:Approximation_quality_convergence}
\end{figure}

For $m<m_h$, i.e., outside of the convergence radius of the hopping expansion, we observe that our models significantly outperform both the hopping expansion and GMRES, and this effect remains when applying a model that was trained with gauge fields generated from a different action (see \cref{app:Action_dependence}).
This is a first indication that the models learn novel expansions of the quark propagator beyond the hopping expansion and polynomials of $\mD$.
To investigate this further, we consider the dependence of the approximation quality on the number of layers of the model.
For $m\gtrsim-0.5$ the test score no longer improves reliably beyond 6 layers, which we attribute to the many possibilities of good low-order approximations in this regime.
For $m\lesssim-0.5$ the test score improves up to the 16 layers we have investigated.
In \cref{fig:Approximation_quality_convergence} we observe that for large mass parameter, all approximations show geometric convergence.
For small mass parameter, the models show geometric convergence up to $16$ layers, while the hopping expansion clearly diverges and GMRES stagnates.
\Cref{app:Convergence_rate} investigates this in more detail and shows that the hopping expansion fails to converge for $m<-1.2$ and GMRES stagnates for $m\lesssim-3$, while the models show geometric convergence for all masses considered.

To summarize, we observe that both model architectures outperform the hopping expansion and GMRES significantly.
In the regime $m\lesssim-0.5$, the HC models perform best and show signs of finding alternative expansions that converge where the hopping expansion does not and GMRES stagnates.
In the regime $m\gtrsim-0.5$, both models show similar performance and learn improved truncations of the hopping expansion.
In the following, we analyze the formulas obtained in each regime.
We restrict ourselves to the HC model for small $m$ and to the restricted model for large $m$.

\subsection{Low-order approximations and novel expansions}\label{sec:Formulas}
By analyzing the model weights as described in \cref{sec:Weight_analysis}, we can extract what the models learned.
In the resulting \cref{eq:Weight_analysis}, we first focus on selected $c_{p,k}$ to get an overview of what the models learned, before providing the learned formulas.

\begin{figure}[ht]
	\centering
	\includegraphics[height=24ex]{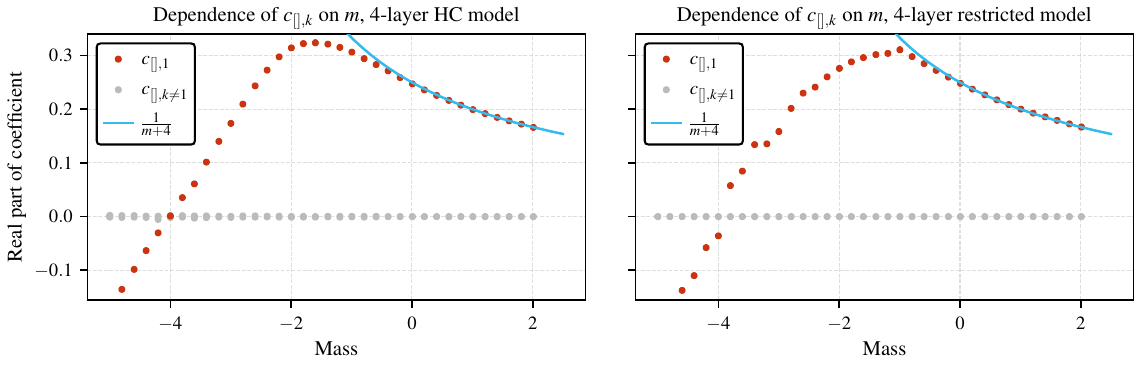}
	\vspace{-1ex}\caption{Dependence of $\textrm{Re}(c_{[],k})$ on the mass parameter, for $4$-layer HC models (left) and $4$-layer restricted models (right).
		The blue line shows the expected $\textrm{Re}(c_{[],1})$ according to the hopping expansion, while the symbols show $\textrm{Re}(c_{[],k})$ as learned by the models.}
	\label{fig:Mass_dependence_coefficients}
\end{figure}

We consider the mass dependence of $c_{p,k}$ for a path $p$ with a fixed path length.
We first investigate the coefficient of the path $[]$ with length 0, i.e., applying no hop operator to the input field.
In the hopping expansion we have $c_{[],1}=2\kappa=\frac1{m+4}$ and $c_{[],k}=0$ for $k\ne1$.
\Cref{fig:Mass_dependence_coefficients} shows the dependence of $\textrm{Re}(c_{[],k})$ on the mass parameter for $4$-layer HC and restricted models, and for the hopping expansion.
$\textrm{Im}(c_{[],k})$ is zero within statistics.
For $m\gtrsim-0.5$, the models clearly recover the hopping expansion, while for $m\lesssim-0.5$, they systematically deviate from it.
Similar behavior can be seen for models with different numbers of layers, other paths, and other $\mGamma_k$, see \cref{app:Mass_dependence}.

We now investigate the low-order approximations obtained in the regime where the hopping expansion is valid.
In \cref{sec:Approximation_quality}, we saw that in this regime it is sufficient to consider the restricted models.
For mass parameter $m=1$, the $n$-th order hopping expansion is
\begin{align}
	D^{-1}=0.2  \sum_{\ell=0}^n\sum_{\mu=1}^4\big((0.1\mI_4+0.1\gamma_\mu)\mH_{+\mu}
	+(0.1\mI_4-0.1\gamma_\mu)\mH_{-\mu}\big)^\ell + \mathcal O(\mH^{n+1})\,.
\end{align}

From the restricted models, we obtain low-order approximations of the quark propagator which improve upon the hopping expansion, such as\footnote{Coefficients are averaged over 5 training runs using different seeds. The standard deviation in the last digits is reported in parentheses.}
\begin{align}
	D^{-1}\approx0.193386(29) \Big(\mI+ & \sum_{\mu=1}^4(0.078651(22)\mI_4+0.08896(8)\gamma_\mu)\mH_{+\mu}  \notag \\[-2.5ex]
	                                    & \qquad +(0.078651(22)\mI_4-0.08896(8)\gamma_\mu)\mH_{-\mu}\Big)\,.
\end{align}

We only show the learned formula for the 1-layer model.
Formulas for deeper models or for different masses are given in \cref{app:Formulas}.
From the standard deviations, we observe that the formulas differ from the hopping expansion in a statistically significant way, and we have already seen in \cref{sec:Approximation_quality} that they lead to a better approximation quality than the hopping expansion or GMRES.
These formulas can be implemented in any framework, providing an improvement in approximation quality in any application where a truncated hopping expansion is currently used.

\begin{figure}[ht]
	\centering
	\includegraphics[height=24ex]{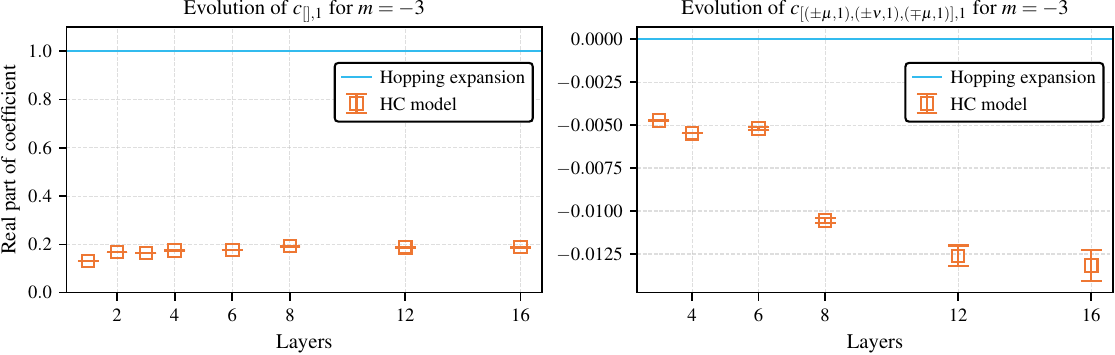}
	\vspace{-1ex}\caption{Evolution of the coefficients $c_{[],1}$ (left) and $c_{[(\pm\mu,1),(\pm\nu,1),(\mp\mu,1)],1}$ (right) at $m=-3$ for HC models, in the latter case averaged over all $\pm\mu$ and $\pm\nu$.}
	\label{fig:Coefficient_evolution}
\end{figure}

Finally, we explore the structure of the approximations obtained by HC models in the regime where the hopping expansion diverges.
By writing the HC models and the hopping expansion in the form of \cref{eq:Weight_analysis}, we can compare the coefficients $c_{p,k}$.
\Cref{fig:Coefficient_evolution} shows two of these coefficients and how they change with the number of layers.
The left plot shows the coefficient $c_{[],1}$, which differs significantly ($0.1882(23)$ for 16 layers) from the hopping-expansion value of $1$ while being very stable as a function of the number of layers.
In the right plot, we show the coefficient $c_{[(\pm\mu,1),(\pm\nu,1),(\mp\mu,1)],1}$ of the staple\footnote{It is called staple because of the shape of the path. The coefficient is identical within statistics for all combinations of $\pm\mu$ and $\pm\nu$. In \cref{fig:Coefficient_evolution} we have averaged over all such combinations.} as an example of a coefficient that is zero in the hopping expansion but nonzero in the expansion found by our model.
The coefficient again stabilizes when increasing the number of layers, this time reaching $-0.0132(9)$ for 16 layers.
The fact that the staple coefficient is nonzero shows that the HC models have learned a structurally new formula.
Plots for other coefficients and other mass parameters are shown in \cref{app:Coefficient_evolution}.
There, we encounter a case in which the HC models learn a coefficient to be zero which was nonzero in the hopping expansion.

\section{Discussion}
Our models outperform the hopping expansion.
They also generalize to unseen gauge fields and to different lattice volumes.
This is made possible by realizing that the hopping expansion is overly restrictive and more freedom can be introduced without additional computational cost.
In the regime where the hopping expansion converges, this freedom is used by our models to systematically change the coefficients of the paths in order to obtain better approximations of the quark propagators.
Outside of this regime, at masses $m\lesssim-0.5$, the models exploit this freedom even more, showing signs of converging to a previously unknown series expansion of the quark propagator.
The convergence regime of this new expansion seems to cover at least the mass-parameter regime $-5<m\lesssim-0.5$.
For larger masses, the models recover the hopping expansion, but with modified coefficients that give a better approximation quality at the same order.
\Cref{app:Mass_dependence} and \cref{fig:Mass_dependence_coefficients} suggest that analytic formulas for the mass dependence of the coefficients can be found, but this needs further work.

\section{Conclusion}
In this work, we used machine-learning methods to improve upon existing approximations of the inverse of a large sparse linear operator, the discretized Dirac operator.
We found evidence of a previously unknown series expansion in the parameter region where the known expansion does not converge.
This demonstrates that machine learning can lead to both algorithmic improvements as well as new insights into well-studied problems in theoretical physics and applied mathematics.

\subsection*{AI use statement}
In this work, we used generative AI tools to help with the code implementation of the method and the result-generating pipeline.
We have not used generative AI tools for any other tasks with required disclosure.
Additionally, we used generative AI tools to analyze existing literature and to help in polishing the paper's writing.
For checking for correctness of presentation and implementation, we used an AI agent to reproduce the results of the paper, given only a paper draft as a reference.
We have reviewed and tested all code written by AI, and we have performed our own
literature analysis independently of AI analysis.
We take responsibility for the final content of this work, including all published code and references produced with the aid of generative AI.

\subsection*{Reproducibility statement}
All gauge fields used for training, as well as all training and analysis scripts, are available at \href{https://github.com/simon-pfahler/low-order-propagators}{github.com/simon-pfahler/low-order-propagators}.
The repository also includes a pipeline to produce the figures used in the paper and generate the required data automatically.

\subsubsection*{Author Contributions}
S.P.\ conceived and designed the project. S.P.\ and M.K.\ implemented the methods and carried out the experiments and analysis. S.P., M.K., D.K.\ and C.L.\ validated the results. S.P.\ and D.K.\ acquired the data. S.P.\ and T.W.\ wrote the manuscript. All authors reviewed and improved the manuscript.

\subsubsection*{Acknowledgments}
This work is funded in part by the German Academic Scholarship Foundation.
It is also funded in part by the German Research Foundation (DFG) through the grant ``PUNCH4NFDI --- Teilchen, Universum, Kerne und Hadronen für die NFDI'' (project 460248186).

\bibliography{references}
\bibliographystyle{iclr2027_conference}

\appendix
\crefalias{section}{appendix}
\section{Approximation Error via Expectation Values}\label{app:Approximation_quality}
The Frobenius norm can be written as an expectation value over random normally distributed and normalized quark fields $\rvv$,
\begin{equation}
	\|\mA\|_F^2=N\cdot\E\left(\|\mA\rvv\|^2\right).
\end{equation}
To derive equation \cref{eq:Approximation_quality}, we compute
\begin{align}
	\E\left(\|\mM\mD\rvv-\rvv\|^2\right)\notag
	 & =\E\left((\mM\mD\rvv-\rvv)^\dag(\mM\mD\rvv-\rvv)\right)                            \\\notag
	 & =\E\left(\rvv^\dag(\mM\mD-\mI)^\dag(\mM\mD-\mI)\rvv\right)                         \\\notag
	 & =\E\left(\textrm{tr}\left((\mM\mD-\mI)^\dag(\mM\mD-\mI)\rvv\rvv^\dag\right)\right) \\\notag
	 & =\textrm{tr}\left((\mM\mD-\mI)^\dag(\mM\mD-\mI)\E(\rvv\rvv^\dag)\right)            \\\notag
	 & =\textrm{tr}\left((\mM\mD-\mI)^\dag(\mM\mD-\mI)\right)\frac1N                      \\
	 & =\frac{\|\mM\mD-\mI\|_F^2}{\|\mI\|_F^2}\,.
\end{align}
In the penultimate step, we have used that $\rvv$ is normalized, i.e., the elements of $\rvv$ are drawn from a complex normal distribution of mean $0$ and variance $1/N$.

\section{Effect of Clifford-Weight Parametrization}\label{app:Clifford_vs_4x4}
We investigate the effect of the weight parametrization in the Clifford layer of \cref{eq:Clifford_layer} on the training speed.
For comparison, we also train models with the linear layer (L) defined in \citet{Pfahler:2026fpj},
\begin{equation}
	\vb_i(\vx)\overset{\textrm{L}}=\sum_{j=1}^n\mW_{ij}\va_j(\vx)\,,
\end{equation}
where $\mW\in\sC^{4\times4}$ are the trainable layer weights.
Both the L and C layer have the same expressivity, as they just parametrize the complex $4\times4$ matrix differently.

\begin{figure}[ht]
	\centering
	\includegraphics[width=\textwidth]{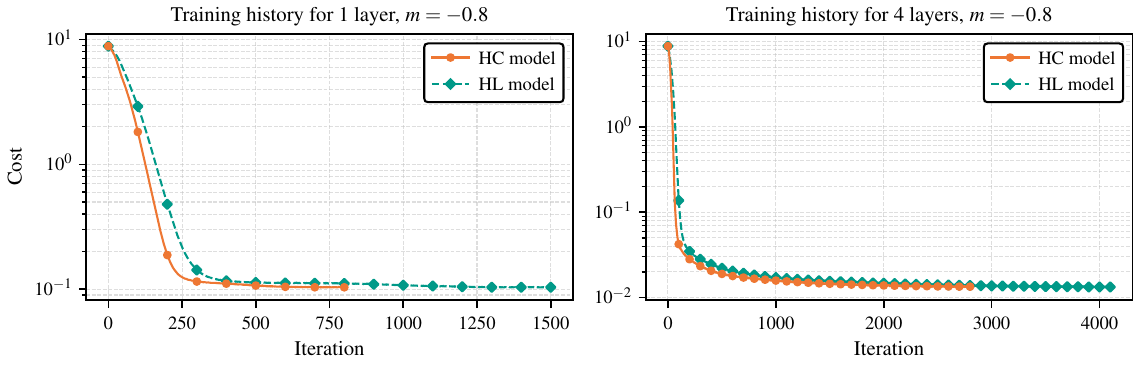}
	\vspace{-1ex}\caption{Training (lines) and test (symbols) scores during the training process of HC and HL models.}
	\label{fig:Clifford_vs_4x4}
\end{figure}

To test their training behavior, we train HC models with $1$ and $4$ layers, and compare them to HL-models, which are HC models with the C layers replaced by L layers.
Training is done as described in \cref{sec:Training_setup}.
In \cref{fig:Clifford_vs_4x4}, we can see that both model types reach the same final score, but the HC models converge faster.

\section{Action Dependence}\label{app:Action_dependence}
We can use different actions for the underlying probability distributions of our gauge fields.
In this appendix, we compare results obtained for three different distributions: Quenched gauge fields generated with the Wilson gauge action at $\beta=6$, dynamical gauge fields generated with the Wilson gauge action at $\beta=5.3$ and two flavors of dynamical fermions with the Wilson-clover Dirac operator at $c_\textrm{SW}=1$ and $m=-0.5$, and random Haar-distributed gauge fields corresponding to $\beta=0$.

\begin{figure}[ht]
	\centering
	\includegraphics[width=\textwidth]{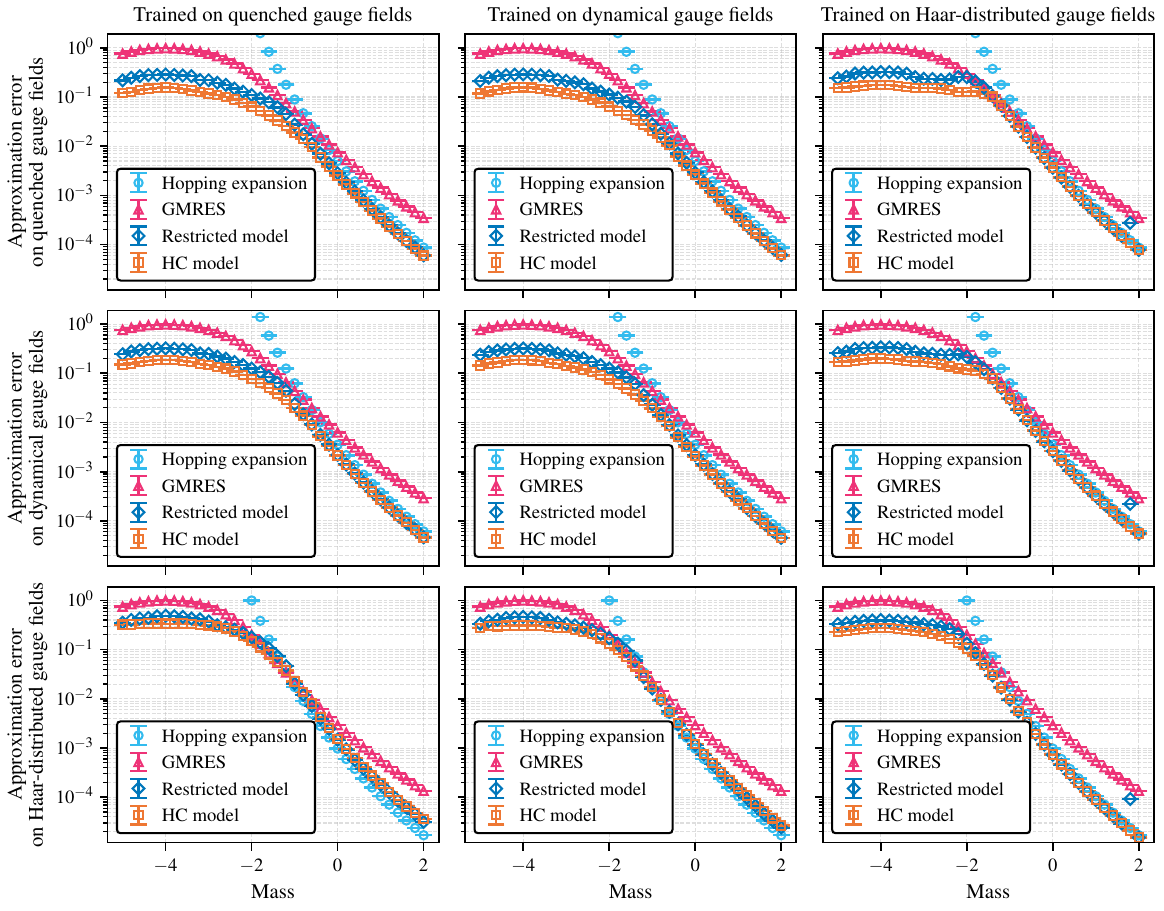}
	\vspace{-1ex}\caption{Approximation error of 4-layer HC and restricted models trained on quenched gauge fields (left column), dynamical gauge fields (middle column), and random Haar-distributed gauge fields (right column), applied on quenched (upper row), dynamical (middle row), and random Haar-distributed (lower row) gauge fields. The hopping expansion and GMRES only differ for different rows.}
	\label{fig:Action_dependence}
\end{figure}

\Cref{fig:Action_dependence} shows that the underlying probability distribution does matter and models perform best when trained and applied on gauge fields with the same probability distribution.
However, only a slight performance degradation is observed applying a model trained on one type of gauge configurations to another type.
Especially in the small-$m$ regime, all models still outperform both the hopping expansion and GMRES even when applied on gauge fields from a different distribution.

\section{Volume Dependence}\label{app:Volume_dependence}
\begin{figure}[ht]
	\centering
	\includegraphics[height=24ex]{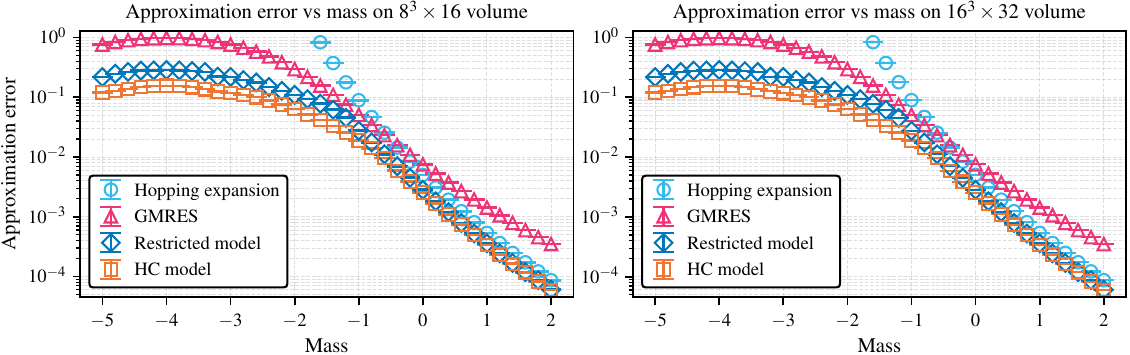}
	\vspace{-1ex}\caption{Approximation error of 4-layer HC and restricted models, compared to the hopping expansion and GMRES, for an $8^3\times16$ (left) and a $16^3\times32$ (right) lattice.}
	\label{fig:Volume_dependence}
\end{figure}

When calculating the approximation error of our models on different lattice volumes, we observe that our models generalize trivially to a different volume than they are trained on, see \cref{fig:Volume_dependence}.
This generalization is expected since the model is structurally similar to a convolutional neural network.

The only noticeable difference in the data is that on the larger volume, the errors are typically smaller.
As mentioned before, this is because the volume provides additional statistics, so the approximation error has a smaller error bar with the same number of samples to calculate \cref{eq:Approximation_quality}.
The effect is barely visible in the plots, but apparent upon inspection of the numbers.

\section{Rate of Convergence}\label{app:Convergence_rate}
\begin{figure}[ht]
	\centering
	\includegraphics[width=0.8\textwidth]{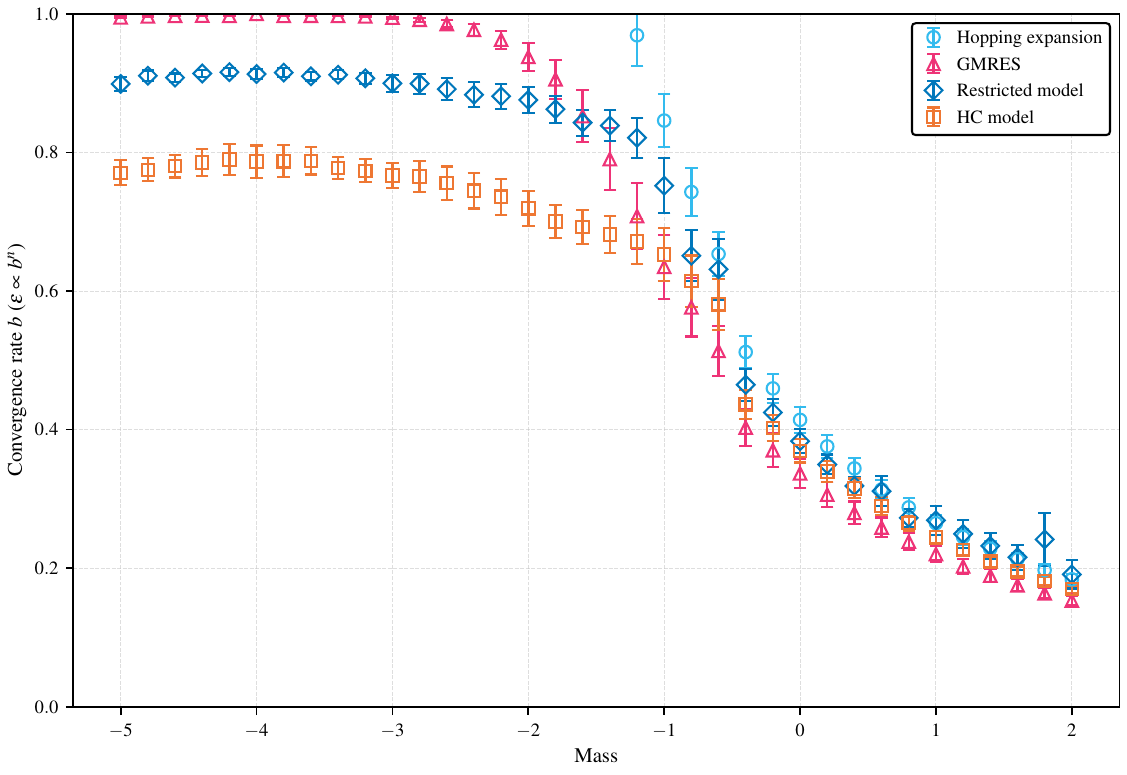}
	\vspace{-1ex}\caption{Rate of convergence of the model approximations, the hopping expansion, and GMRES as a function of the mass parameter.}
	\label{fig:Convergence_rate}
\end{figure}

The hopping expansion is a geometric series, i.e., the approximation error scales with the order $n$ of the expansion (denoted by $\textrm{HE}_n$) as
\begin{equation}
	\varepsilon(\textrm{HE}_n,\mD)=\frac{\|\textrm{HE}_n\mD-\mI\|_F^2}{\|\mI\|_F^2}=\frac{\kappa^{2n+2}\|\mH^{n+1}\|_F^2}{12V}\,.
\end{equation}
To study the convergence behavior, we fit the $n$-dependence of $\varepsilon$ to a curve $a\cdot b^n$, where the parameter $b$ is the rate of convergence.
\Cref{fig:Convergence_rate} shows this rate of convergence for the hopping expansion, the HC and the restricted models, using data from models with 1, 2, 3, 4, and 6 layers for each mass-parameter value.
For $m<-0.5$, models with 12 and 16 layers were additionally used for the fit.
We observe that the hopping expansion diverges (i.e., $b\ge1$) for $m\lesssim-1.2$ and GMRES stagnates (i.e., $b\approx1$) for $m\lesssim-3$, while the trained models converge (i.e., $b<1$) up to the smallest masses tested.

We also observe that while our models provide improved low-order approximations for all masses (see \cref{sec:Approximation_quality}), the convergence of GMRES is faster for $m\gtrsim-1$, since it can adapt to the quark field $\va$ for which we approximate $\mD^{-1}\va$.
For $m\lesssim-1$, our models show faster convergence than any of the baselines, adding to the evidence that they approach a previously unknown convergent expansion.

\section{Mass Dependence of Coefficients}\label{app:Mass_dependence}
\begin{figure}[ht]
	\centering
	\includegraphics[width=0.87\textwidth]{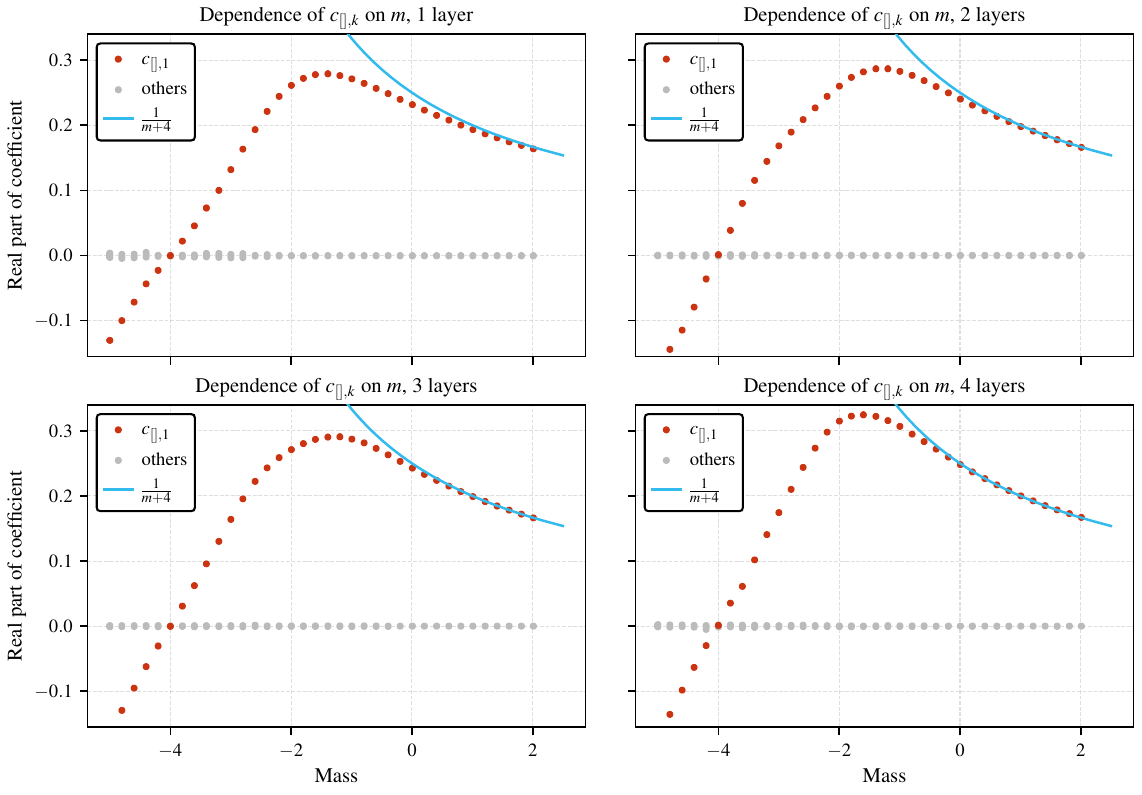}
	\vspace{-1ex}\caption{Dependence of the real part of coefficients of length-zero paths on the mass for HC models with different numbers of layers.}
	\label{fig:coefficients_0hop}
\end{figure}

\begin{figure}[ht]
	\centering
	\includegraphics[width=0.87\textwidth]{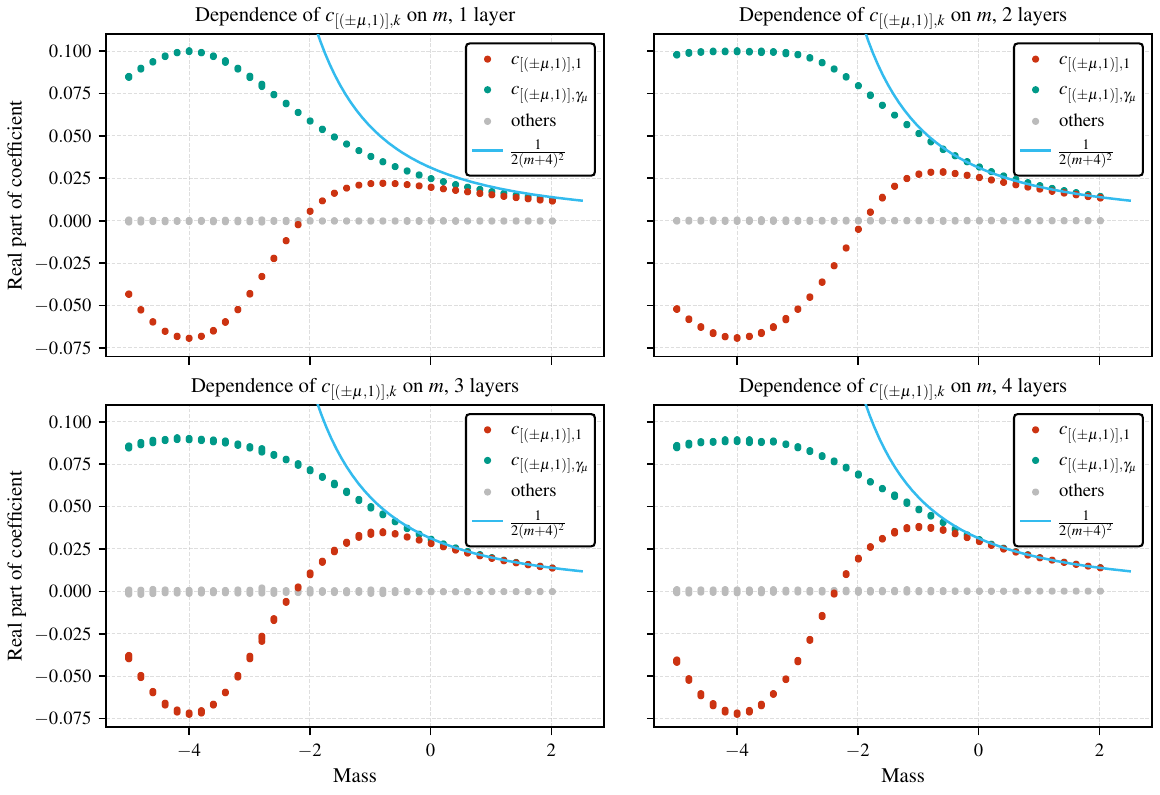}
	\vspace{-1ex}\caption{Same as \cref{fig:coefficients_0hop}, but for paths of length one.}
	\label{fig:coefficients_1hop}
\end{figure}

\begin{figure}[ht]
	\centering
	\includegraphics[width=0.9\textwidth]{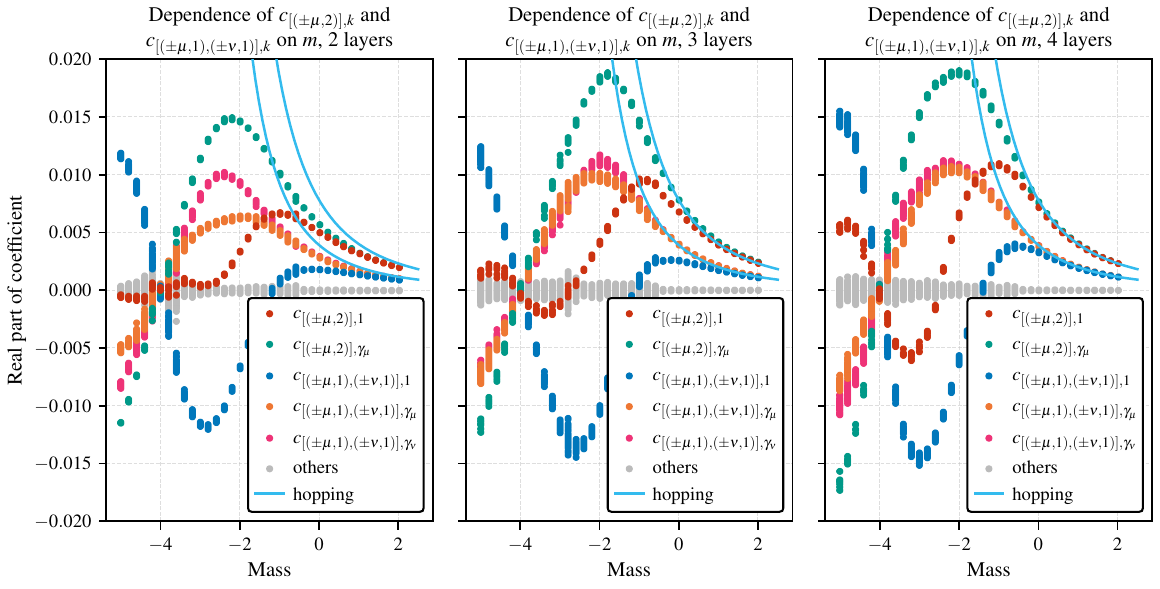}
	\vspace{-1ex}\caption{Same as \cref{fig:coefficients_0hop}, but for paths of length two.}
	\label{fig:coefficients_2hop}
\end{figure}

We show additional results for the mass dependence of the coefficients obtained from model weights, for HC models only.
Restricted models show similar behavior, but the results are somewhat less smooth in the small-$m$ regime where the restricted models do not perform well.
\Cref{fig:coefficients_0hop,fig:coefficients_1hop,fig:coefficients_2hop} show the mass dependence of the coefficients $c_{p,k}$ for $p$ with length zero, one, and two, respectively.
In all plots, it can be seen that the coefficients change smoothly as a function of the mass, indicating that analytic formulas for the mass dependence of the coefficients can be found.

\section{Improved Low-Order Approximation Formulas}\label{app:Formulas}
We provide formulas for improved low-order approximations obtained from the restricted models for mass parameters $m=0,1,2$ and $n=1,2,3,4,6$ layers.
All formulas have the form
\begin{align}
	D^{-1}=w\sum_{\ell=0}^n\prod_{\ell'=1}^\ell\sum_{\mu=1}^4\big((w_\mI^{(\ell')}\mI_4+w_\gamma^{(\ell')}\gamma_\mu)\mH_{+\mu}+(w_\mI^{(\ell')}\mI_4-w_\gamma^{(\ell')}\gamma_\mu)\mH_{-\mu}\big)+\mathcal{O}(\mH^{n+1})\,,
\end{align}
and we provide the coefficients $w$, $w_\mI^{(\ell)}$, and $w_\gamma^{(\ell)}$ in \cref{table:Formulas_m0,table:Formulas_m1,table:Formulas_m2}.
For other masses, these formulas can be obtained in a straightforward manner using the code provided as supplementary material.

\begin{table}[ht]
	\centering
	\vspace{-1ex}\caption{Coefficients of the restricted models for mass parameter $m=0$ and $n$ layers. The standard deviation in the last digits is reported in parentheses. The parenthesized numbers in the header row indicate the value of the weights in the hopping expansion. The imaginary parts of the weights are not displayed if they are consistent with zero within statistics.}
	\label{table:Formulas_m0}
	\begin{tabular}{ccd{1.20}d{1.17}d{1.18}}
		\toprule
		$n$ & $\ell$ & \multicolumn{1}{c}{$w$ (0.25)} & \multicolumn{1}{c}{$w_\mI^{(\ell)}$ (0.125)} & \multicolumn{1}{c}{$w_\gamma^{(\ell)}$ (0.125)} \\
		\midrule
		1   & 1      & 0.231536(19)                   & 0.085127(16)                                 & 0.10732(7)                                      \\
		\midrule
		2   & 1      & 0.25081(5)                     & 0.09250(7)                                   & 0.11711(5)                                      \\
		    & 2      &                                & 0.09267(7)                                   & 0.11688(6)                                      \\
		\midrule
		3   & 1      & 0.2542(11)                     & 0.1017(12)                                   & 0.114(9)                                        \\
		    & 2      &                                & 0.1014(13)                                   & 0.122(7)                                        \\
		    & 3      &                                & 0.1022(9)                                    & 0.121(9)                                        \\
		\midrule
		4   & 1      & 0.25239(21)-0.0009(7)i         & 0.106(4)                                     & 0.122(9)                                        \\
		    & 2      &                                & 0.1094(27)                                   & 0.127(7)                                        \\
		    & 3      &                                & 0.107(4)+0.0035(14)i                         & 0.123(8)-0.0040(16)i                            \\
		    & 4      &                                & 0.107(4)                                     & 0.121(10)                                       \\
		\midrule
		6   & 1      & 0.2507(4)                      & 0.1067(25)                                   & 0.128(14)                                       \\
		    & 2      &                                & 0.109(11)                                    & 0.131(5)                                        \\
		    & 3      &                                & 0.115(6)                                     & 0.132(13)                                       \\
		    & 4      &                                & 0.1181(14)                                   & 0.138(7)                                        \\
		    & 5      &                                & 0.109(11)                                    & 0.1360(24)                                      \\
		    & 6      &                                & 0.111(8)                                     & 0.134(4)                                        \\
		\bottomrule
	\end{tabular}
\end{table}

\begin{table}[ht]
	\centering
	\vspace{-1ex}\caption{Same as \cref{table:Formulas_m0}, but for $m=1$.}
	\label{table:Formulas_m1}
	\begin{tabular}{ccd{1.10}d{1.20}d{1.20}}
		\toprule
		$n$ & $\ell$ & \multicolumn{1}{c}{$w$ (0.2)} & \multicolumn{1}{c}{$w_\mI^{(\ell)}$ (0.1)} & \multicolumn{1}{c}{$w_\gamma^{(\ell)}$ (0.1)} \\
		\midrule
		1   & 1      & 0.193386(29)                  & 0.078651(22)                               & 0.08896(8)                                    \\
		\midrule
		2   & 1      & 0.200858(6)                   & 0.084093(16)                               & 0.09601(5)                                    \\
		    & 2      &                               & 0.084121(23)                               & 0.095910(27)                                  \\
		\midrule
		3   & 1      & 0.2011(5)                     & 0.0908(17)+0.00025(18)i                    & 0.098(5)-0.00027(20)i                         \\
		    & 2      &                               & 0.0888(18)                                 & 0.094(6)                                      \\
		    & 3      &                               & 0.0907(16)                                 & 0.098(5)                                      \\
		\midrule
		4   & 1      & 0.20045(5)                    & 0.0909(8)                                  & 0.0936(13)                                    \\
		    & 2      &                               & 0.0966(9)                                  & 0.1026(8)                                     \\
		    & 3      &                               & 0.092(4)                                   & 0.099(4)                                      \\
		    & 4      &                               & 0.0937(24)                                 & 0.100(6)                                      \\
		\midrule
		6   & 1      & 0.20017(11)                   & 0.082(13)                                  & 0.085(17)                                     \\
		    & 2      &                               & 0.092(10)                                  & 0.094(13)                                     \\
		    & 3      &                               & 0.081(13)                                  & 0.083(17)                                     \\
		    & 4      &                               & 0.089(15)                                  & 0.094(15)                                     \\
		    & 5      &                               & 0.087(13)                                  & 0.093(13)                                     \\
		    & 6      &                               & 0.091(14)                                  & 0.094(16)                                     \\
		\bottomrule
	\end{tabular}
\end{table}

\begin{table}[ht]
	\centering
	\vspace{-1ex}\caption{Same as \cref{table:Formulas_m0}, but for $m=2$.}
	\label{table:Formulas_m2}
	\begin{tabular}{ccd{1.10}d{1.16}d{1.16}}
		\toprule
		$n$ & $\ell$ & \multicolumn{1}{c}{$w$ (0.166\ldots)} & \multicolumn{1}{c}{$w_\mI^{(\ell)}$ (0.0833\ldots)} & \multicolumn{1}{c}{$w_\gamma^{(\ell)}$ (0.0833\ldots)} \\
		\midrule
		1   & 1      & 0.163918(21)                          & 0.070676(16)                                        & 0.07619(6)                                             \\
		\midrule
		2   & 1      & 0.1671247(30)                         & 0.074569(30)                                        & 0.080976(9)                                            \\
		    & 2      &                                       & 0.074544(21)                                        & 0.080985(16)                                           \\
		\midrule
		3   & 1      & 0.16711(20)                           & 0.0785(19)                                          & 0.080(4)                                               \\
		    & 2      &                                       & 0.0776(11)                                          & 0.079(5)                                               \\
		    & 3      &                                       & 0.0793(6)                                           & 0.0835(7)                                              \\
		\midrule
		4   & 1      & 0.16684(7)                            & 0.074(16)                                           & 0.073(14)                                              \\
		    & 2      &                                       & 0.076(15)                                           & 0.076(14)                                              \\
		    & 3      &                                       & 0.074(17)                                           & 0.075(15)                                              \\
		    & 4      &                                       & 0.075(14)                                           & 0.075(14)                                              \\
		\midrule
		6   & 1      & 0.166697(29)                          & 0.074(5)                                            & 0.076(5)                                               \\
		    & 2      &                                       & 0.073(13)                                           & 0.075(14)                                              \\
		    & 3      &                                       & 0.055(11)-0.0006(5)i                                & 0.055(11)+0.0006(5)i                                   \\
		    & 4      &                                       & 0.061(18)                                           & 0.064(19)                                              \\
		    & 5      &                                       & 0.068(18)                                           & 0.071(17)                                              \\
		    & 6      &                                       & 0.073(5)                                            & 0.075(6)                                               \\
		\bottomrule
	\end{tabular}
\end{table}

\section{Coefficient Evolution}\label{app:Coefficient_evolution}
\begin{figure}[ht]
	\centering
	\includegraphics[width=0.8\textwidth]{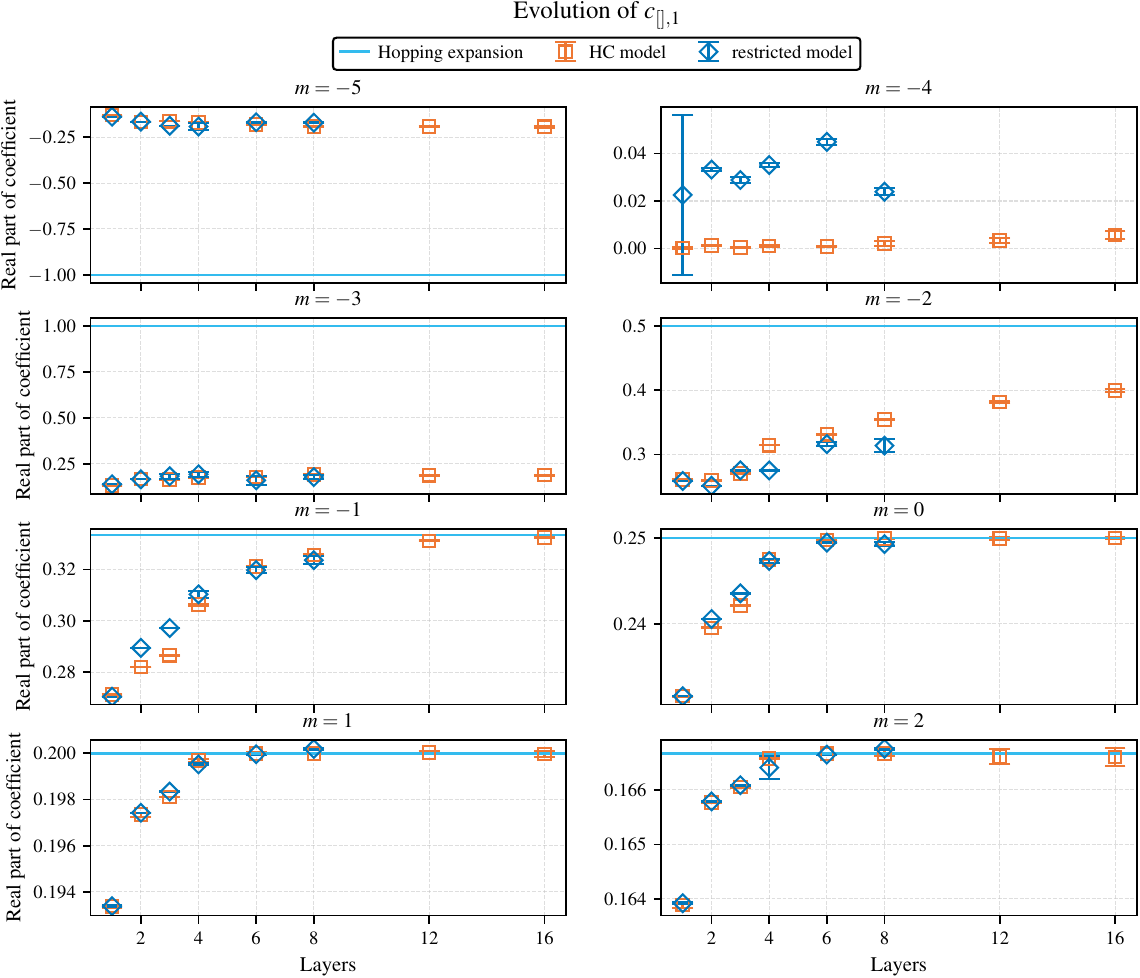}
	\vspace{-1ex}\caption{Evolution of the coefficient $c_{[],1}$ as a function of the number of layers for HC models at different masses.}
	\label{fig:evolution_nohop}
\end{figure}

\begin{figure}[ht]
	\centering
	\includegraphics[width=0.8\textwidth]{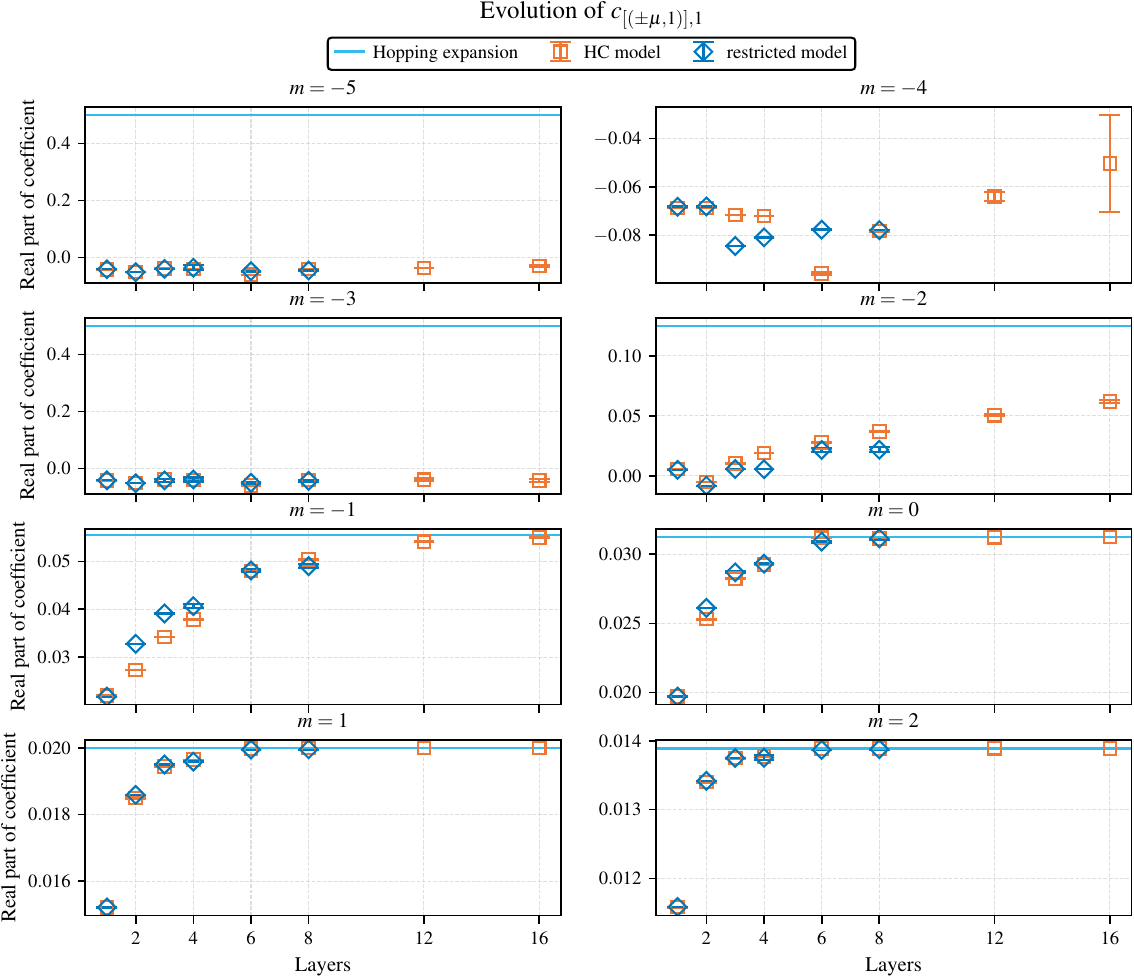}
	\vspace{-1ex}\caption{Same as \cref{fig:evolution_nohop}, but for $c_{[(\pm\mu,1)],1}$, averaged over $\pm\mu$.}
	\label{fig:evolution_1hop_id}
\end{figure}

\begin{figure}[ht]
	\centering
	\includegraphics[width=0.8\textwidth]{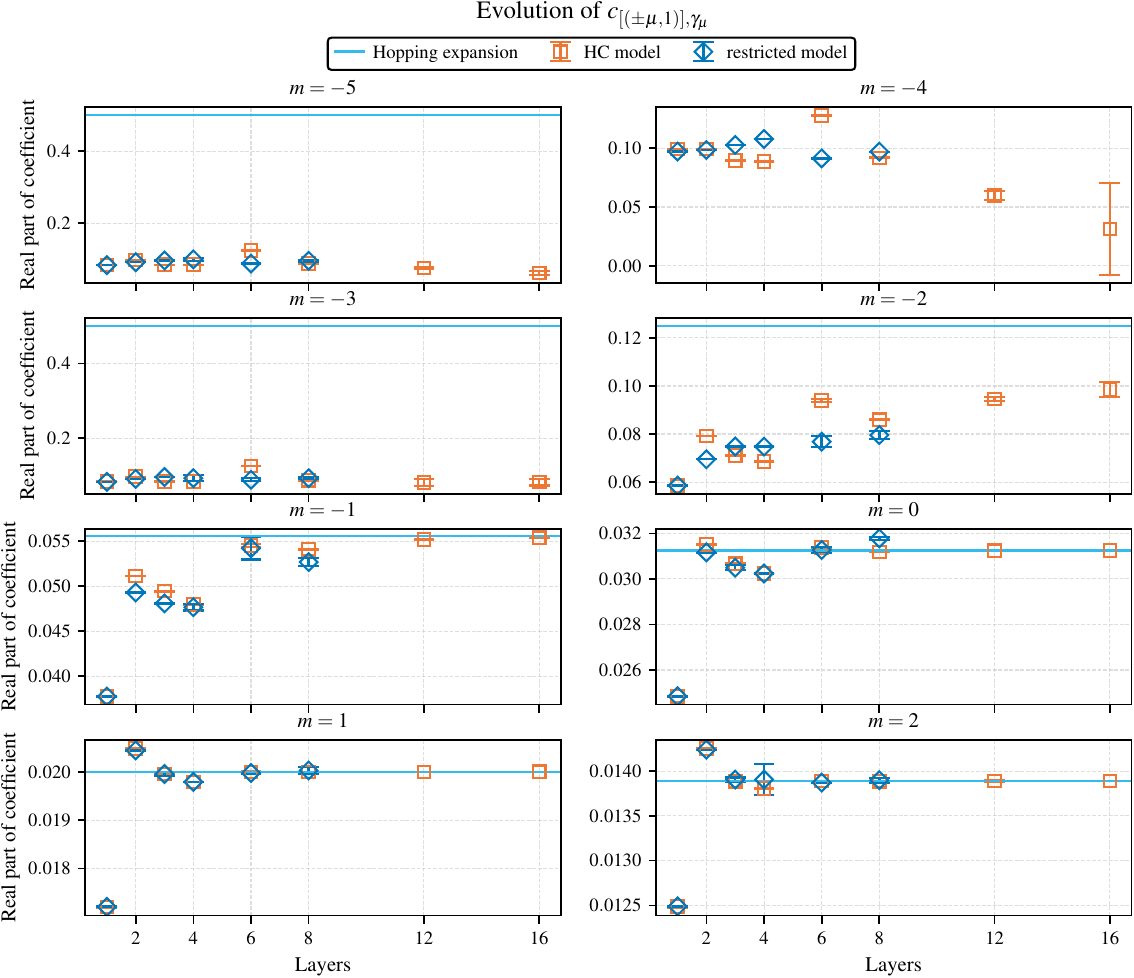}
	\vspace{-1ex}\caption{Same as \cref{fig:evolution_nohop}, but for $c_{[(\pm\mu,1)],\gamma_\mu}$, averaged over $\pm\mu$.}
	\label{fig:evolution_1hop_gamma}
\end{figure}

\begin{figure}[ht]
	\centering
	\includegraphics[width=0.8\textwidth]{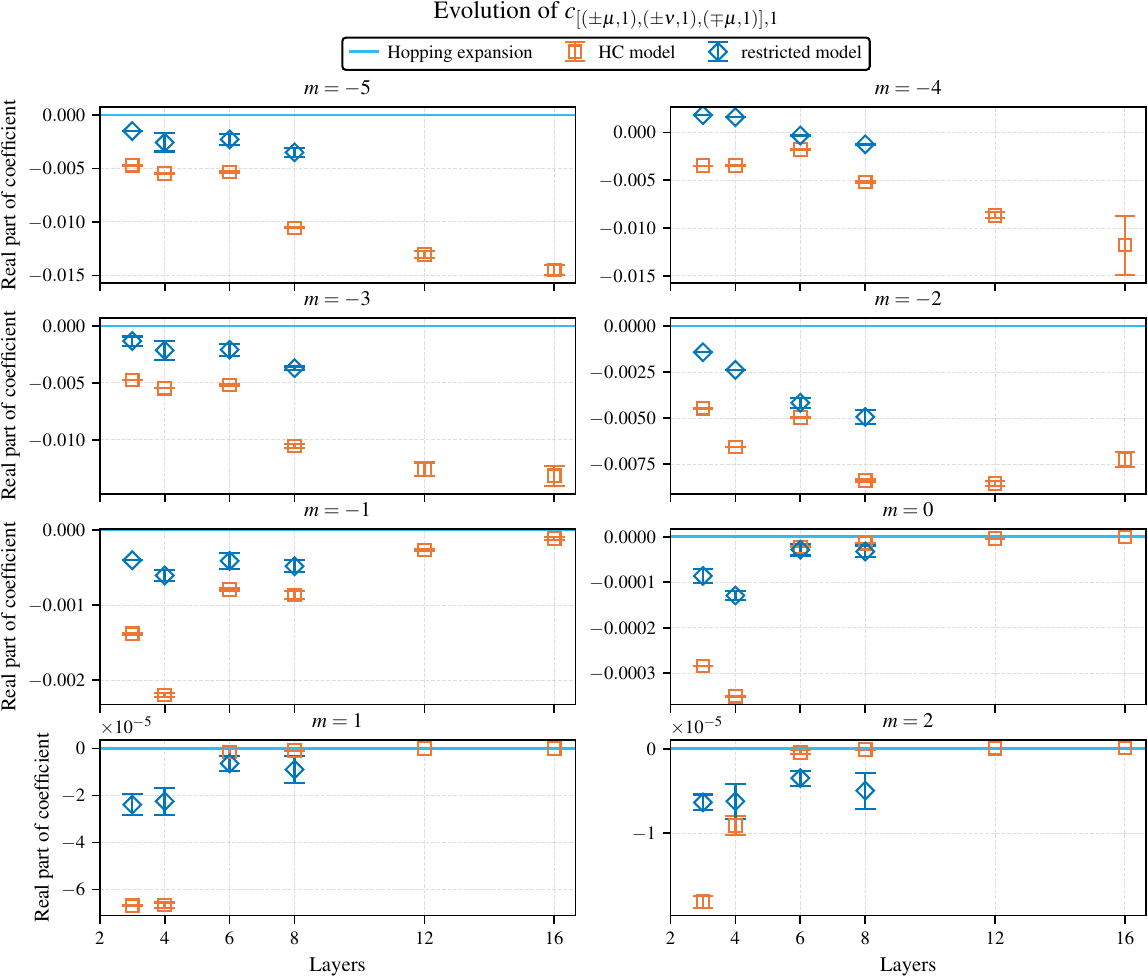}
	\vspace{-1ex}\caption{Same as \cref{fig:evolution_nohop}, but for $c_{[(\pm\mu,1),(\pm\nu,1),(\mp\mu,1)],1}$, averaged over $\pm\mu$ and $\pm\nu$.}
	\label{fig:evolution_staple_id}
\end{figure}

\begin{figure}[ht]
	\centering
	\includegraphics[width=0.8\textwidth]{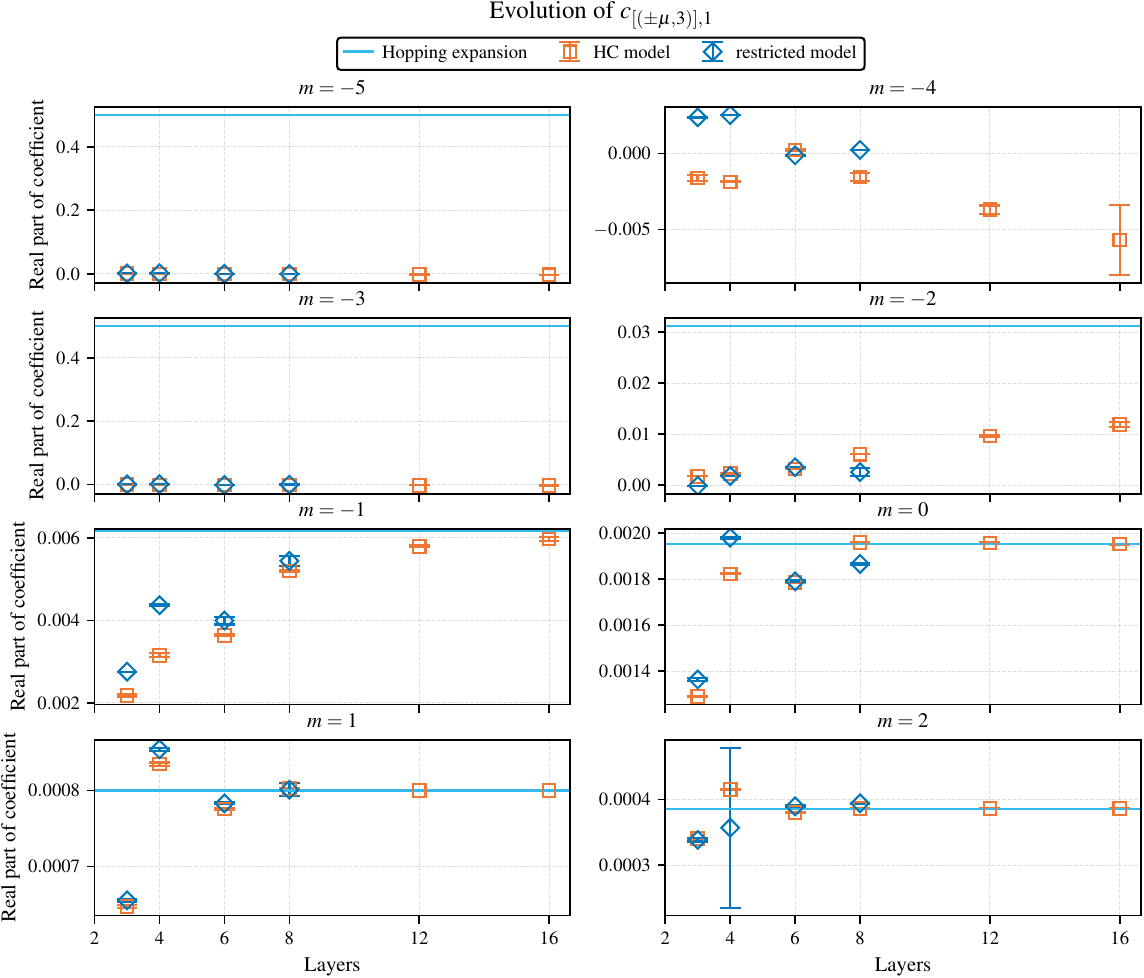}
	\vspace{-1ex}\caption{Same as \cref{fig:evolution_nohop}, but for $c_{[(\pm\mu,3)],1}$, averaged over $\pm\mu$.}
	\label{fig:evolution_3shop}
\end{figure}

We show the evolution of the coefficients $c_{p,k}$ in \cref{eq:Weight_analysis} for HC and restricted models as a function of the number of layers, up to 8 layers.
The coefficients $c_{[],1}$ are shown in \cref{fig:evolution_nohop}, and the coefficients $c_{[(\pm\mu,1)],1}$ and $c_{[(\pm\mu,1)],\gamma_\mu}$ are shown in \cref{fig:evolution_1hop_id,fig:evolution_1hop_gamma}.
As an example for a coefficient that is zero in the hopping expansion but nonzero for our models in the small-$m$ regime, we show the staple coefficient $c_{[(\pm\mu,1),(\pm\nu,1),(\mp\mu,1)],1}$ in \cref{fig:evolution_staple_id}.
Finally, in \cref{fig:evolution_3shop} we show an example for a coefficient that is nonzero in the hopping expansion, but zero for our models in the small-$m$ regime, i.e., $c_{[(\pm\mu,3)],1}$.

\end{document}

%% file: math_commands.tex
\usepackage{amsmath,amsfonts,bm}

\def\eqref#1{equation~\ref{#1}}

\def\1{\bm{1}}

\def\rvv{{\mathbf{v}}}

\def\va{{\bm{a}}}
\def\vb{{\bm{b}}}

\def\ve{{\bm{e}}}

\def\vx{{\bm{x}}}
\def\vy{{\bm{y}}}

\def\mA{{\bm{A}}}

\def\mD{{\bm{D}}}

\def\mH{{\bm{H}}}
\def\mI{{\bm{I}}}

\def\mM{{\bm{M}}}

\def\mT{{\bm{T}}}
\def\mU{{\bm{U}}}

\def\mW{{\bm{W}}}

\DeclareMathAlphabet{\mathsfit}{\encodingdefault}{\sfdefault}{m}{sl}
\SetMathAlphabet{\mathsfit}{bold}{\encodingdefault}{\sfdefault}{bx}{n}

\def\sC{{\mathbb{C}}}

\def\sN{{\mathbb{N}}}

\def\sR{{\mathbb{R}}}

\newcommand{\E}{\mathbb{E}}



%% file: main.bbl
\begin{thebibliography}{40}
\providecommand{\natexlab}[1]{#1}
\providecommand{\url}[1]{\texttt{#1}}
\expandafter\ifx\csname urlstyle\endcsname\relax
  \providecommand{\doi}[1]{doi: #1}\else
  \providecommand{\doi}{doi: \begingroup \urlstyle{rm}\Url}\fi

\bibitem[Abbott et~al.(2026)Abbott, Boyda, Fu, Hackett, Kanwar,
  {Romero-L{\'o}pez}, Shanahan, and Urban]{Abbott:2026ylv}
Ryan Abbott, Denis Boyda, Yang Fu, Daniel~C. Hackett, Gurtej Kanwar, Fernando
  {Romero-L{\'o}pez}, Phiala~E. Shanahan, and Julian~M. Urban.
\newblock Variance reduction in lattice {{QCD}} observables via normalizing
  flows.
\newblock \emph{arXiv Preprint}, March 2026.
\newblock \doi{10.48550/arXiv.2603.02984}.

\bibitem[Aliberti et~al.(2025)]{Aliberti:2025bega}
R.~Aliberti et~al.
\newblock The anomalous magnetic moment of the muon in the {{Standard Model}}:
  An update.
\newblock \emph{Physics Reports}, 1143:\penalty0 1--158, 2025.
\newblock \doi{10.1016/j.physrep.2025.08.002}.

\bibitem[Aoki et~al.(2026)Aoki, Blum, Collins, Debbio, Morte, Dimopoulos, Feng,
  Golterman, Gottlieb, Gupta, Herdo{\'i}za, Hernandez, J{\"u}ttner, Kaneko,
  Lunghi, Meinel, Monahan, Nicholson, Onogi, Petreczky, Portelli, Ramos,
  Sharpe, Simone, Sint, Sommer, Tantalo, de~Water, Vaquero, Wenger, and
  Wittig]{FlavourLatticeAveragingGroupFLAG:2024oxs}
Y.~Aoki, T.~Blum, S.~Collins, L.~Del Debbio, M.~Della Morte, P.~Dimopoulos,
  X.~Feng, M.~Golterman, Steven Gottlieb, R.~Gupta, G.~Herdo{\'i}za,
  P.~Hernandez, A.~J{\"u}ttner, T.~Kaneko, E.~Lunghi, S.~Meinel, C.~Monahan,
  A.~Nicholson, T.~Onogi, P.~Petreczky, A.~Portelli, A.~Ramos, S.~R. Sharpe,
  J.~N. Simone, S.~Sint, R.~Sommer, N.~Tantalo, R.~Van de~Water, A.~Vaquero,
  U.~Wenger, and H.~Wittig.
\newblock {{FLAG Review}} 2024.
\newblock \emph{Physical Review D}, 113\penalty0 (1):\penalty0 014508, January
  2026.
\newblock ISSN 2470-0010, 2470-0029.
\newblock \doi{10.1103/nfzp-p5dn}.

\bibitem[Barbour et~al.(1988)Barbour, Davies, and Sabeur]{Barbour:1988ax}
I.~M. Barbour, C.~T.~H. Davies, and Z.~Sabeur.
\newblock Lattice {{QCD}} at finite density.
\newblock \emph{Physics Letters B}, 215\penalty0 (3):\penalty0 567--572,
  December 1988.
\newblock ISSN 0370-2693.
\newblock \doi{10.1016/0370-2693(88)91361-5}.

\bibitem[Bhuyan et~al.(2024)Bhuyan, {Ramdane-Cherif}, Tomar, and
  Singh]{Bhuyan_2024}
Bikram~Pratim Bhuyan, Amar {Ramdane-Cherif}, Ravi Tomar, and T.~P. Singh.
\newblock Neuro-symbolic artificial intelligence: A survey.
\newblock \emph{Neural Computing and Applications}, 36\penalty0 (21):\penalty0
  12809--12844, July 2024.
\newblock ISSN 1433-3058.
\newblock \doi{10.1007/s00521-024-09960-z}.

\bibitem[Biggio et~al.(2020)Biggio, Bendinelli, Lucchi, and
  Parascandolo]{Biggio_2020}
Luca Biggio, Tommaso Bendinelli, Aurelien Lucchi, and Giambattista
  Parascandolo.
\newblock A {{Seq2Seq}} approach to {{Symbolic Regression}}.
\newblock In \emph{Learning {{Meets Combinatorial Algorithms}} at
  {{NeurIPS2020}}}, November 2020.

\bibitem[Bornyakov(2017)]{Bornyakov:2016wld}
V.~G. Bornyakov.
\newblock New approach to canonical partition functions computation in {{Nf}}=2
  lattice {{QCD}} at finite baryon density.
\newblock \emph{Physical Review D}, 95\penalty0 (9), 2017.
\newblock \doi{10.1103/PhysRevD.95.094506}.

\bibitem[Boyda et~al.(2022)Boyda, Cal{\`i}, Foreman, Funcke, Hackett, Lin,
  Aarts, Alexandru, Jin, Lucini, and Shanahan]{Boyda:2022nmh}
Denis Boyda, Salvatore Cal{\`i}, Sam Foreman, Lena Funcke, Daniel~C. Hackett,
  Yin Lin, Gert Aarts, Andrei Alexandru, Xiao-Yong Jin, Biagio Lucini, and
  Phiala~E. Shanahan.
\newblock Applications of {{Machine Learning}} to {{Lattice Quantum Field
  Theory}}.
\newblock \emph{arXiv Preprint}, February 2022.
\newblock \doi{10.48550/arXiv.2202.05838}.

\bibitem[Bronstein et~al.(2021)Bronstein, Bruna, Cohen, and Veli{\v
  c}kovi{\'c}]{Bronstein:2021mdi}
Michael~M. Bronstein, Joan Bruna, Taco Cohen, and Petar Veli{\v c}kovi{\'c}.
\newblock Geometric {{Deep Learning}}: {{Grids}}, {{Groups}}, {{Graphs}},
  {{Geodesics}}, and {{Gauges}}.
\newblock \emph{arXiv Preprint}, May 2021.
\newblock \doi{10.48550/arXiv.2104.13478}.

\bibitem[Camps et~al.(2025)Camps, Rrapaj, Klymko, Kim, Gott, Darbha, Balewski,
  Austin, and Wright]{Camps:2025adg}
Daan Camps, Ermal Rrapaj, Katherine Klymko, Hyeongjin Kim, Kevin Gott, Siva
  Darbha, Jan Balewski, Brian Austin, and Nicholas~J. Wright.
\newblock Quantum {{Computing Technology Roadmaps}} and {{Capability
  Assessment}} for {{Scientific Computing}} -- {{An}} analysis of use cases
  from the {{NERSC}} workload.
\newblock \emph{arXiv Preprint}, September 2025.
\newblock \doi{10.48550/arXiv.2509.09882}.

\bibitem[Cohen \& Welling(2016)Cohen and Welling]{Cohen_2016}
Taco Cohen and Max Welling.
\newblock Group {{Equivariant Convolutional Networks}}.
\newblock In \emph{Proceedings of {{The}} 33rd {{International Conference}} on
  {{Machine Learning}}}, pp.\  2990--2999. PMLR, June 2016.

\bibitem[Dong \& Zhong(2025)Dong and Zhong]{Dong_2025}
Junlan Dong and Jinghui Zhong.
\newblock Recent {{Advances}} in {{Symbolic Regression}}.
\newblock \emph{ACM Computing Surveys}, 57\penalty0 (11):\penalty0
  290:1--290:37, June 2025.
\newblock ISSN 0360-0300.
\newblock \doi{10.1145/3735634}.

\bibitem[Favoni et~al.(2022)Favoni, Ipp, M{\"u}ller, and Schuh]{Favoni:2020reg}
Matteo Favoni, Andreas Ipp, David~I. M{\"u}ller, and Daniel Schuh.
\newblock Lattice {{Gauge Equivariant Convolutional Neural Networks}}.
\newblock \emph{Physical Review Letters}, 128\penalty0 (3):\penalty0 032003,
  January 2022.
\newblock \doi{10.1103/PhysRevLett.128.032003}.

\bibitem[Goy et~al.(2017)Goy, Bornyakov, Boyda, Molochkov, Nakamura, Nikolaev,
  and Zakharov]{Goy:2016egl}
V.~A. Goy, V.~Bornyakov, D.~Boyda, A.~Molochkov, A.~Nakamura, A.~Nikolaev, and
  V.~Zakharov.
\newblock Sign problem in finite density lattice {{QCD}}.
\newblock \emph{Progress of Theoretical and Experimental Physics},
  2017\penalty0 (3):\penalty0 031D01, March 2017.
\newblock ISSN 2050-3911.
\newblock \doi{10.1093/ptep/ptx018}.

\bibitem[Hasenbusch(2018)]{Hasenbusch:2018iuw}
Martin Hasenbusch.
\newblock Exploiting the hopping parameter expansion in the hybrid {{Monte
  Carlo}} simulation of lattice {{QCD}} with two degenerate flavors of
  {{Wilson}} fermions.
\newblock \emph{Physical Review D}, 97\penalty0 (11):\penalty0 114512, June
  2018.
\newblock \doi{10.1103/PhysRevD.97.114512}.

\bibitem[Hasenfratz \& Hasenfratz(1981)Hasenfratz and
  Hasenfratz]{Hasenfratz:1981nc}
Anna Hasenfratz and Peter Hasenfratz.
\newblock Monte {{Carlo}} simulation combined with a hopping parameter
  expansion.
\newblock \emph{Physics Letters B}, 104\penalty0 (6):\penalty0 489--493,
  September 1981.
\newblock ISSN 0370-2693.
\newblock \doi{10.1016/0370-2693(81)90521-9}.

\bibitem[Kanwar et~al.(2020)Kanwar, Albergo, Boyda, Cranmer, Hackett,
  Racani{\`e}re, Rezende, and Shanahan]{Kanwar:2020xzo}
Gurtej Kanwar, Michael~S. Albergo, Denis Boyda, Kyle Cranmer, Daniel~C.
  Hackett, S{\'e}bastien Racani{\`e}re, Danilo~Jimenez Rezende, and Phiala~E.
  Shanahan.
\newblock Equivariant {{Flow-Based Sampling}} for {{Lattice Gauge Theory}}.
\newblock \emph{Physical Review Letters}, 125\penalty0 (12):\penalty0 121601,
  September 2020.
\newblock \doi{10.1103/PhysRevLett.125.121601}.

\bibitem[Kingma \& Ba(2017)Kingma and Ba]{Kingma:2014vow}
Diederik~P. Kingma and Jimmy Ba.
\newblock Adam: {{A Method}} for {{Stochastic Optimization}}.
\newblock \emph{arXiv Preprint}, January 2017.
\newblock \doi{10.48550/arXiv.1412.6980}.

\bibitem[Kitazawa \& Wada(2026)Kitazawa and Wada]{Kitazawa:2026vbv}
Masakiyo Kitazawa and Tatsuya Wada.
\newblock Higher-order hopping-parameter expansion by human-{{AI}}
  collaboration.
\newblock \emph{arXiv Preprint}, August 2026.
\newblock \doi{10.48550/arXiv.2606.31492}.

\bibitem[Kn{\"u}ttel(2025{\natexlab{a}})]{Knuttel_2025}
Daniel Kn{\"u}ttel.
\newblock {{ILDG Gauge Field Ensemble}} ldg/rqcd/{{qcdML}}/ens\_001, March
  2025{\natexlab{a}}.
\newblock Dataset. doi: 10.5281/zenodo.15018324.

\bibitem[Kn{\"u}ttel(2025{\natexlab{b}})]{Knuttel_2025a}
Daniel Kn{\"u}ttel.
\newblock {{ILDG Gauge Field Ensemble}}
  mc://ldg/rqcd/{{qcdML}}/ens\_004\_selected, October 2025{\natexlab{b}}.
\newblock Dataset. doi: 10.5281/zenodo.17494830.

\bibitem[Kronfeld(2012)]{Kronfeld:2012uk}
Andreas~S. Kronfeld.
\newblock Twenty-{{First Century Lattice Gauge Theory}}: {{Results}} from the
  {{Quantum Chromodynamics Lagrangian}}.
\newblock \emph{Annual Review of Nuclear and Particle Science}, 62:\penalty0
  265--284, 2012.
\newblock ISSN 1545-4134.
\newblock \doi{10.1146/annurev-nucl-102711-094942}.

\bibitem[Langelage \& Philipsen(2010)Langelage and Philipsen]{Langelage:2010yn}
Jens Langelage and Owe Philipsen.
\newblock The pressure of strong coupling lattice {{QCD}} with heavy quarks,
  the hadron resonance gas and the large {{N}} limit.
\newblock \emph{Journal of High Energy Physics}, 2010\penalty0 (4):\penalty0
  55, April 2010.
\newblock ISSN 1029-8479.
\newblock \doi{10.1007/JHEP04(2010)055}.

\bibitem[Lehner \& Wettig(2023)Lehner and Wettig]{Lehner:2023bba}
C.~Lehner and T.~Wettig.
\newblock Gauge-equivariant neural networks as preconditioners in lattice
  {{QCD}}.
\newblock \emph{Physical Review D}, 108\penalty0 (3):\penalty0 034503, August
  2023.
\newblock \doi{10.1103/PhysRevD.108.034503}.

\bibitem[Lehner \& Wettig(2024)Lehner and Wettig]{Lehner:2023prf}
C.~Lehner and T.~Wettig.
\newblock Gauge-equivariant pooling layers for preconditioners in lattice
  {{QCD}}.
\newblock \emph{Physical Review D}, 110\penalty0 (3):\penalty0 034517, August
  2024.
\newblock \doi{10.1103/PhysRevD.110.034517}.

\bibitem[Martius \& Lampert(2016)Martius and Lampert]{Martius_2016}
Georg Martius and Christoph~H. Lampert.
\newblock Extrapolation and learning equations.
\newblock \emph{arXiv Preprint}, October 2016.
\newblock \doi{10.48550/arXiv.1610.02995}.

\bibitem[Mundhenk et~al.(2021)Mundhenk, Landajuela, Glatt, Santiago, Faissol,
  and Petersen]{Mundhenk_2021}
T.~Nathan Mundhenk, Mikel Landajuela, Ruben Glatt, Claudio~P. Santiago,
  Daniel~M. Faissol, and Brenden~K. Petersen.
\newblock Symbolic {{Regression}} via {{Neural-Guided Genetic Programming
  Population Seeding}}.
\newblock \emph{arXiv Preprint}, November 2021.
\newblock \doi{10.48550/arXiv.2111.00053}.

\bibitem[Nakamura et~al.(2016)Nakamura, Oka, and Taniguchi]{Nakamura:2015jra}
Atsushi Nakamura, Shotaro Oka, and Yusuke Taniguchi.
\newblock {{QCD}} phase transition at real chemical potential with canonical
  approach.
\newblock \emph{Journal of High Energy Physics}, 2016\penalty0 (2):\penalty0
  54, February 2016.
\newblock ISSN 1029-8479.
\newblock \doi{10.1007/JHEP02(2016)054}.

\bibitem[{NPLQCD Collaboration} et~al.(2024){NPLQCD Collaboration}, Davoudi,
  Detmold, Fu, Grebe, Jay, Murphy, Oare, Shanahan, and Wagman]{Davoudi:2024ukx}
{NPLQCD Collaboration}, Zohreh Davoudi, William Detmold, Zhenghao Fu,
  Anthony~V. Grebe, William Jay, David Murphy, Patrick Oare, Phiala~E.
  Shanahan, and Michael~L. Wagman.
\newblock Long-distance nuclear matrix elements for neutrinoless double-beta
  decay from lattice {{QCD}}.
\newblock \emph{Physical Review D}, 109\penalty0 (11):\penalty0 114514, June
  2024.
\newblock \doi{10.1103/PhysRevD.109.114514}.

\bibitem[Petersen et~al.(2021)Petersen, Landajuela, Mundhenk, Santiago, Kim,
  and Kim]{Petersen_2021}
Brenden~K. Petersen, Mikel Landajuela, T.~Nathan Mundhenk, Claudio~P. Santiago,
  Soo~K. Kim, and Joanne~T. Kim.
\newblock Deep symbolic regression: {{Recovering}} mathematical expressions
  from data via risk-seeking policy gradients.
\newblock \emph{arXiv Preprint}, April 2021.
\newblock \doi{10.48550/arXiv.1912.04871}.

\bibitem[Pfahler et~al.(2026)Pfahler, Kn{\"u}ttel, Lehner, and
  Wettig]{Pfahler:2026fpj}
Simon Pfahler, Daniel Kn{\"u}ttel, Christoph Lehner, and Tilo Wettig.
\newblock A novel gauge-equivariant neural network architecture for
  preconditioners in lattice {{QCD}}.
\newblock In \emph{Proceedings of {{The}} 42nd {{International Symposium}} on
  {{Lattice Field Theory}} --- {{PoS}}({{LATTICE2025}})}, volume 518, pp.\
  027. SISSA Medialab, April 2026.
\newblock \doi{10.22323/1.518.0027}.

\bibitem[Philipsen(2021)]{Philipsen:2021vgp}
Owe Philipsen.
\newblock Strong coupling methods in {{QCD}} thermodynamics.
\newblock \emph{Indian Journal of Physics}, 95\penalty0 (8):\penalty0
  1599--1611, August 2021.
\newblock ISSN 0974-9845.
\newblock \doi{10.1007/s12648-021-02164-4}.

\bibitem[Saad(2003)]{Saad_2003}
Y.~Saad.
\newblock \emph{Iterative {{Methods}} for {{Sparse Linear Systems}}}.
\newblock {Society for Industrial and Applied Mathematics}, April 2003.
\newblock ISBN 978-0-89871-534-7.

\bibitem[Schmidt \& Lipson(2009)Schmidt and Lipson]{Schmidt_2009}
Michael Schmidt and Hod Lipson.
\newblock Distilling {{Free-Form Natural Laws}} from {{Experimental Data}}.
\newblock \emph{Science}, 324\penalty0 (5923):\penalty0 81--85, April 2009.
\newblock \doi{10.1126/science.1165893}.

\bibitem[Suarez et~al.(2025)Suarez, Amaya, Frank, Freyermuth, Girone,
  Kostrzewa, and Pfalzner]{Suarez:2025cnd}
Estela Suarez, Jorge Amaya, Martin Frank, Oliver Freyermuth, Maria Girone,
  Bartosz Kostrzewa, and Susanne Pfalzner.
\newblock Energy efficiency trends in {{HPC}}: What high-energy and
  astrophysicists need to know.
\newblock \emph{Frontiers in Physics}, 13, April 2025.
\newblock ISSN 2296-424X.
\newblock \doi{10.3389/fphy.2025.1542474}.

\bibitem[Udrescu \& Tegmark(2020)Udrescu and Tegmark]{Udrescu:2019mnk}
Silviu-Marian Udrescu and Max Tegmark.
\newblock {{AI Feynman}}: {{A}} physics-inspired method for symbolic
  regression.
\newblock \emph{Science Advances}, 6\penalty0 (16):\penalty0 eaay2631, April
  2020.
\newblock ISSN 2375-2548.
\newblock \doi{10.1126/sciadv.aay2631}.

\bibitem[Wakabayashi et~al.(2022)Wakabayashi, Ejiri, Kanaya, and
  Kitazawa]{Wakabayashi:2021eye}
Naoki Wakabayashi, Shinji Ejiri, Kazuyuki Kanaya, and Masakiyo Kitazawa.
\newblock Scope and convergence of the hopping parameter expansion in
  finite-temperature quantum chromodynamics with heavy quarks around the
  critical point.
\newblock \emph{Progress of Theoretical and Experimental Physics},
  2022\penalty0 (3):\penalty0 033B05, March 2022.
\newblock ISSN 2050-3911.
\newblock \doi{10.1093/ptep/ptac019}.

\bibitem[Wetzel et~al.(2025)Wetzel, Ha, Iten, Klopotek, and
  Liu]{Wetzel:2025uhj}
Sebastian~Johann Wetzel, Seungwoong Ha, Raban Iten, Miriam Klopotek, and Ziming
  Liu.
\newblock Interpretable {{Machine Learning}} in {{Physics}}: {{A Review}}.
\newblock \emph{arXiv Preprint}, March 2025.
\newblock \doi{10.48550/arXiv.2503.23616}.

\bibitem[Wilson(1974)]{Wilson:1974sk}
Kenneth~G. Wilson.
\newblock Confinement of quarks.
\newblock \emph{Physical Review D}, 10\penalty0 (8):\penalty0 2445--2459,
  October 1974.
\newblock \doi{10.1103/PhysRevD.10.2445}.

\bibitem[Zhang et~al.(2026)Zhang, Zhang, Hu, Lu, Pang, and Wang]{Zhang:2026bqa}
Wei-Jie Zhang, Zhenyu Zhang, Jifeng Hu, Bing-Nan Lu, Jin-Yi Pang, and Qian
  Wang.
\newblock Machine {{Learning Unveils Finite-volume Energy Shifts}} in
  {{Three-body System}}.
\newblock \emph{arXiv Preprint}, August 2026.
\newblock \doi{10.48550/arXiv.2608.28700}.

\end{thebibliography}
